\documentclass[11pt]{article}
\usepackage{epsfig} 
\usepackage{amssymb} 
\usepackage{a4}

\def\ra{\rightarrow}

\newcommand\nutau{{\nu_\tau}}

\newcommand\numu{{\nu_\mu}}

\newcommand\nue{{\nu_e}}

\def\dm2{\Delta m^2}
\def\sq2{sin^2(2\Theta)}

\def\osc{\rightsquigarrow}
\def\tet23{\theta_{23}}

\begin{document}

\vspace*{1cm}
\begin{center}
{\Large{\bf Neutrino Factories: Detector Concepts for studies \\ of CP and T violation effects
in neutrino oscillations\footnote{Based on an invited talk at the IX International Workshop on ``Neutrino Telescopes'',
March 6-9, 2001, Venice (Italy).} }}\\
\vspace{.5cm}
Andr\'e Rubbia\\
Institut f\"{u}r Teilchenphysik, ETHZ, CH-8093 Z\"{u}rich,
Switzerland
\end{center}

\abstract{
The ideal neutrino detector at the neutrino factory should have a mass in
the range of 10~kton, provide particle identification to tag the flavor of
the incoming neutrino, lepton charge
measurement to select the incoming neutrino helicity, 
good energy resolution to reconstruct the incoming
neutrino energy, and be isotropic to equally well reconstruct incoming
neutrinos from different baselines (it might be more efficient to build various
sources at different baselines, than various detectors). The detector
should also be able to reconstruct neutrino event typically below 15~GeV.
A detector with such quality is most adapted to fully study neutrino oscillations at the
neutrino factory. In particular, measurement of the leading muons and electrons charge is
the only way to fully simultaneously explore $CP$ and $T$ violation
effects. We think that a magnetized liquid argon imaging detector stands
today as the best choice of technique, that holds the highest promises to
match the above mentioned detector requirements. We discuss also the
optimal neutrino factory energy and baseline between source and detector in
order to best perform these studies.}

\section{Introduction}
The first generation long baseline (LBL) experiments ---
K2K~\cite{k2k}, MINOS~\cite{minos}, OPERA~\cite{opera} and
ICARUS~\cite{icarus,Arneodo:2001tx} --- will 
use artificial neutrino beams produced by 
the ``traditional'' meson-decay method, to
search for
a conclusive and unambiguous signature of the neutrino flavor
oscillation~\cite{pontecorvo}
observed in cosmic ray neutrinos~\cite{atmevid}.
These experiments will provide the first precise
measurements of the parameters governing the main
muon disappearance mechanism. 

In contrast, a neutrino factory\cite{geers,nufacwww} is understood 
as a machine where
low energy muons of a given charge are accelerated to high energy,
and let decay into one electron and two neutrinos within a muon storage ring.

The great physics potential of a neutrino factory comes
from its ability to test in a very clean and high statistics
environment all possible flavor oscillation transitions~\cite{nufacfis,Bueno:2000fg}.
This ability will provide very
stringent information on all the elements of the neutrino mixing matrix and
on the mass pattern of the neutrinos. 

In a $3\times 3$-mixing scenario, the mixing matrix, which should be
unitary, 
is determined by three angles and a complex phase. 
Neutrino factories will provide precise determinations of two angles and 
of the largest mass difference\cite{nufacfis,Bueno:2000fg}. In addition, a test of the
unitarity 
of the matrix could be performed\cite{Bueno:2000fg}.

Apart from being able to measure very precisely all the magnitude of the
elements of the mixing matrix, {\it the more challenging and most interesting
goal of the neutrino factory is the search for effects related to the
complex phase of the mixing matrix}\cite{Arafune:1997hd}. 
The complex phase will in general alter
the neutrino flavor oscillation probabilities, and will most strikingly
introduce a difference of transition probabilities between neutrinos and
antineutrinos (CP-violation effects), and between 
time-reversed transitions (T-violation
effects).
Neutrino factories should provide the intense 
and well controlled beams needed to perform these studies.

It should be stressed that the complete and comprehensive detection 
of $T$- and/or $CP-$violation
effects 
is very difficult for terrestrial experiments, as it requires $L/E_\nu$ values simultaneously in the
range of solar and atmospheric neutrinos. In addition, they require that
the
transitions $\nue\ra\numu$, $\bar\nue\ra\bar\numu$, $\numu\ra\nue$ and
$\bar\numu\ra\bar\nue$ be measured, a priori within the same experiment.
It has been known\cite{nufacfis} that only in the case of the LMA solution
to the solar neutrino problem, can one hope to look for effects related to
the complex nature of the mixing matrix. We assume that this is the case.

\begin{figure}[tb]
\centering
\epsfig{file=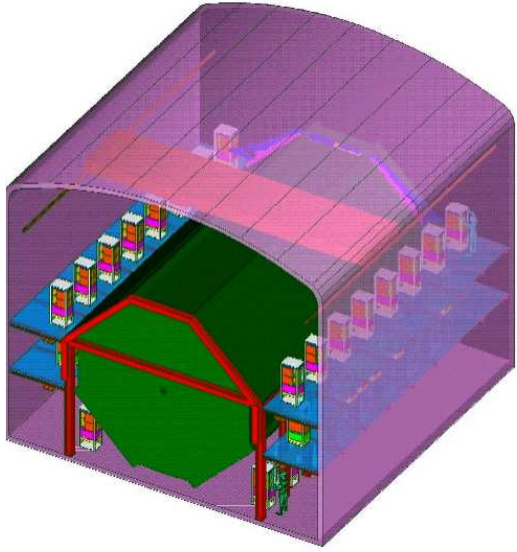,width=7cm}
\caption{View of the planned MINOS detector. The detector with a total mass
of 5.4~kton should start data
taking with the FNAL-NUMI beam in 2004.}
\label{fig:minos}
\end{figure}

\section{Detectors at the neutrino factory}
We briefly mention the kind of detectors that are currently envisaged in
the context of the neutrino factory.

\subsection{Magnetized steel-scintillator sandwich}
This is the ``traditional''
neutrino detector, with a lot of experience gained with (though smaller)
detectors like CCFR/NuTeV or CDHS. A detector based on iron 
has the advantage of having a high density and to be
easily magnetizable, for a ``straight-forward'' $\mu^+/\mu^-$
discrimination. It has sufficient granularity to only cleanly detect muons, 
and offers a rather poor discrimination and reconstruction of electrons and neutral-current-like
events (including hadronic decays of taus). The muon resolution is good and
the jet energy resolution is reasonable (typ. $80\%/\sqrt{E_h}$). 
A minimum muon energy threshold (typ. 4-6 GeV) is needed in order to
separate the muon from other hadrons and the
muon separation from the jet is difficult to measure. The electron/hadron
discrimination is rather poor. The angular resolution is determined by the
transverse readout segmentation, which is in fact rather modest.
The full volume of the detector has to be instrumented, and by nature of
the sandwich, the readout is not very isotropic. Hence, a detector optimized
to reconstruct ``horizontal'' events from an artificial neutrino beam coming from a well defined
direction, will not at the same time provide good reconstruction of say
atmospheric events coming from ``above'' and ``below''.
One considers as
prototype for the neutrino factory
the MINOS\cite{minos} detector (see Figure~\ref{fig:minos}), 
that should reach a mass of 5.4 kton in
2003. It is composed of 486 layers of 2.45cm~Fe each, divided into two
sections each 15~m long. The field has an average value of 1.3~T. Since
detector construction is under way, we shall soon learn if this detector
technology can be scaled to the larger masses currently envisaged for a
neutrino factory, namely in the range of 40~kton.

\subsection{Large water-Cerenkov}
\begin{figure}[htb]
\centering
\epsfig{file=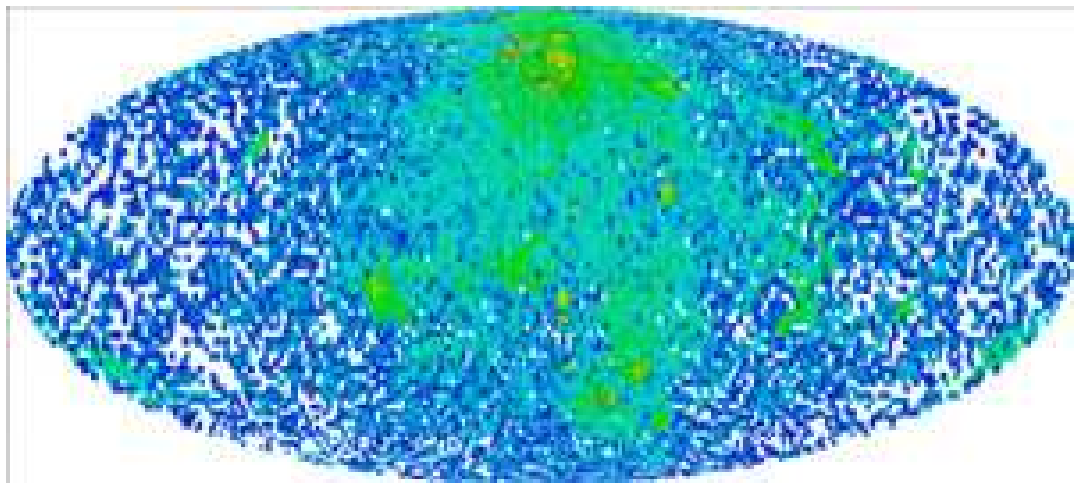,width=14cm}
\caption{Simulated neutrino event from a 50~GeV storage ring.}
\label{fig:sk50gev}
\end{figure}

This is a well-proven technology (IMB,
Kamiokande, SuperKamiokande). The target material (Water) is cheap and only
the surface of the detector (and not the volume unlike in the previous
magnetized
steel-scintillator sandwich) needs to be instrumented,
hence this technology scales well to gigantic masses. The size will
eventually be limited by water properties. A next generation 500~kton is
under consideration\cite{Jung:1999jq}. The reconstruction of events is
rather isotropic, and this kind of detector will certainly cover a broad
physics program, including observation of atmospheric neutrinos, maybe
solar neutrinos if energy threshold allows it, supernova neutrinos, and search
for proton decays.
The clear disadvantage of this kind of detector is that the
reconstructed pattern is limited to ``simple event topologies'', that are
not really compatible with a detailed reconstruction of
neutrino events with complicated final states. 
See Figure~\ref{fig:sk50gev}. The detection of the muon charge cannot be
easily implemented in water and should rely on a downstream muon
identifier.

\subsection{Emulsion/target sandwich}
The ECC (Emulsion Cloud Chamber) technique currently envisaged for
OPERA~\cite{opera} has been used successfully in the DONUT\cite{Sielaff:2001md} experiment,
though in a very small scale compared to what is needed for a neutrino factory.
The technique should be demonstrated at the 1~kton scale at the LNGS by the year
2005. Emulsions have a fantastic granularity (at the level of microns) and 
can be used to directly detect the kink in the decay of a charged tau
lepton. In OPERA, emulsions will be used as very precise tracking devices
to reconstruct pieces of tracks, whereas actual neutrino interactions will
occur in a passive target material (Pb). The non-alignment of the track segments
reconstructed over a thickness of about 100~microns of two consecutive
emulsion layers, will indicate potential decay kinks. 
A direct search for $\nue\ra\nutau$ at a neutrino factory could be attempted if the charge
of the tau can be detected to suppress the $\numu\ra\nutau$
``background''. The most obvious disadvantages are the difficulty to scale
this detector to very large masses, the potentially very large amount of
scanning involved in a high-statistics experiment at a neutrino factory and
possibly a severe background from charm decays produced in $\nue$
interactions. 

\begin{figure}[htb]
\begin{center}
\mbox{\epsfig{file=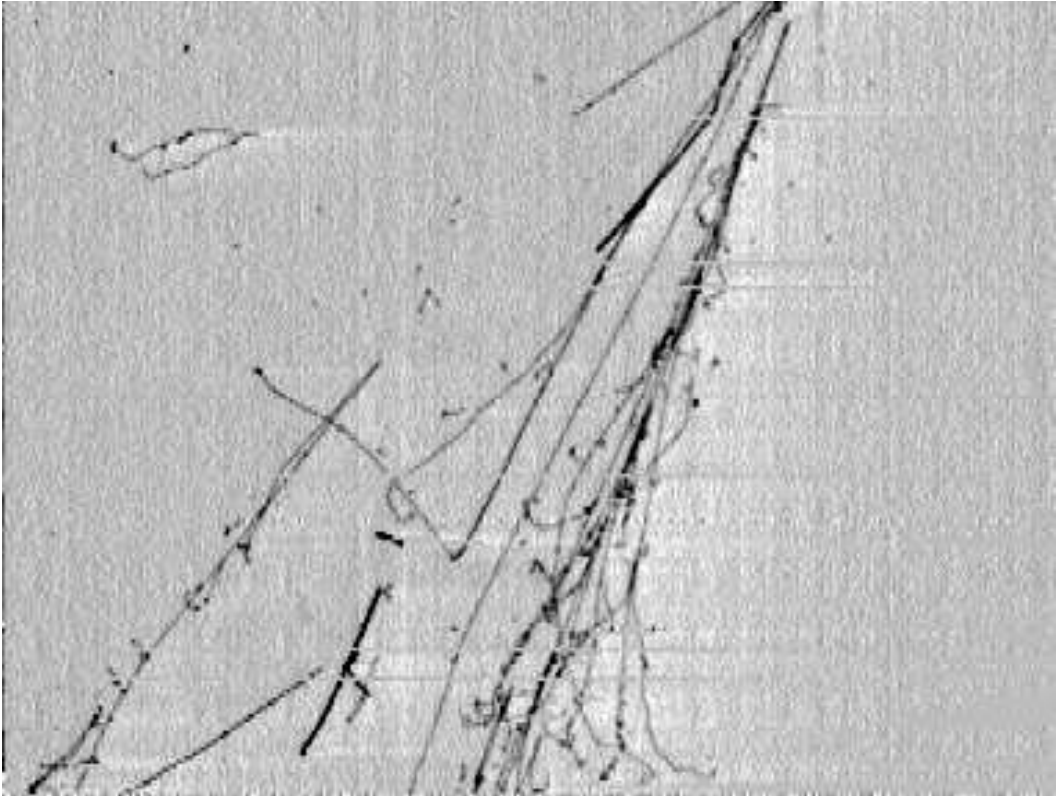,width=14cm}} 
\vspace{-0.5truecm}     
\caption{Electronic liquid argon imaging of a cosmic ray induced shower.  The overall drift time (horizontal 
axis) corresponds to about 40 cm of drift distance.  The vertical co-ordinate is the wire 
numbering; around 40 cm are shown.}
\label{fig:3ton} 
\end{center}
\end{figure}

\begin{figure}[htb]
\begin{center}
\mbox{\epsfig{file=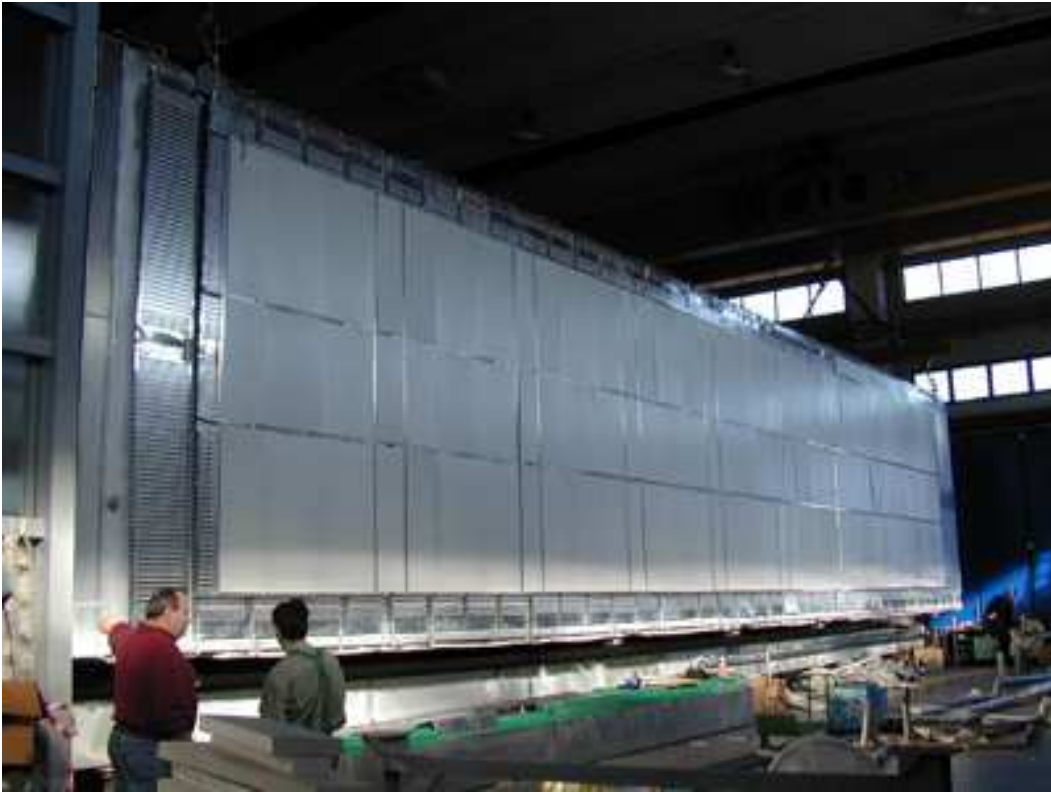,width=14cm}} 
\vspace{-0.5truecm}     
\caption{{\it Side view during installation
of one of the two cryostats that
compose the ICARUS T600 detector.}}
\label{fig:cryo} 
\end{center}
\end{figure}

\subsection{Liquid Argon imaging TPC}
The liquid argon imaging technique provides fully bubble-chamber-like
reconstruction of events (See Figure~\ref{fig:3ton}). It is fully a homogeneous, continuous, precise
tracking device with high resolution $dE/dx$ measurement and full sampling
electromagnetic and hadronic calorimetry. Imaging provides excellent
electron identification and electron/hadron separation. Energy resolution
is excellent (typ. $3\%/\sqrt{E}$ for e.m. showers) and the hadronic energy
resolution of contained events is also excellent
(typ. $30\%/\sqrt{E_h}$).

Like with the water Cerenkov detector, liquid argon detectors cover a broad
physics program, including observation of atmospheric neutrinos, 
solar neutrinos, supernova neutrinos, and search for
proton decays.

One disadvantage in the current implementation is the lack of
magnetic field. One has in principle to rely on a
down-stream muon spectrometer (that has however low threshold given the
loss in Argon of $240~MeV/m$). Magnetization was considered in the past\cite{cr}, and
a possibility to magnetize the large, multikton, volume of Argon is however under
study\cite{sergiam}. This method is the only one, most easily scalable to multikton mass
range, 
that would provide sufficient granularity to measure the charge of
electrons (see Section~5).

Liquid argon imaging, though a priori more difficult to implement than say
a magnetized iron-scintillator sandwich, is becoming a mature
technique, that has so far been demonstrated up to the 15 ton prototype
scale. The next major milestone is the operation of the ICARUS 600 ton
prototype
(see Figure~\ref{fig:cryo}).
Its construction has been completed during 2000, in all 
its various components. The first technical run has started in 
March 2001. Cosmic muon tracks have been seen in June 2001\footnote{See
{\it http://www.cern.ch/icarus/} for up-to-date information}.
The successful reaching of this milestone is very important, and after
a series of other technical tests to be performed in the assembly hall within the
summer 2001, the detector should be ready to be transported to the LNGS
tunnel by middle 2002. 
The physics program achievable with the T600 detector has been
described in Ref.\cite{Arneodo:2001tx}. It covers the observation and study
of atmospheric and solar neutrinos.

\section{The oscillation physics at the neutrino factory}
Neutrino sources from muon decays provide clear advantages over
neutrino beams from pion decays.
The {\it exact neutrino helicity composition} is a fundamental tool to study
neutrino oscillations. It can be easily selected,
since $\mu^+\ra e^+\nue\bar\numu$
and $\mu^-\ra e^-\bar\nue\numu$ can be separately obtained. 

At a neutrino factory,
one could independently study the following flavor transitions:
\begin{eqnarray}
\mu^-\ra e^-& \bar\nue&\numu \nonumber \\ 
& & \ra \nue\ra e^-\rm \ appearance\\
& & \ra \numu\rm \ disappearance, \ same \ sign \ muons\\
& & \ra \nutau\ra\tau^-\rm \ appearance, \ high \ energy \ nu's\\
 & \ra &\bar\nue\rm \ disappearance\\
 & \ra &\bar\numu\ra\mu^+\rm \ appearance, \ wrong \ sign \ muons\\
 & \ra &\bar\nutau\ra\tau^+\rm \ appearance, \ high \ energy \ nu's
\end{eqnarray}
plus 6 other charge conjugate processes initiated
from $\mu^+$ decays.

{\it The ideal neutrino detector should be able to measure these 12
different processes as a function of the baseline $L$ and of the neutrino
energy $E_\nu$!}

Of particular interest are the charged current neutrino interactions, since
they can in principle be used to tag the neutrino flavor and helicity,
through the detection and identification of the final state charged lepton:
\begin{eqnarray}
\nu_\ell N \ra \ell^-+hadrons\ \ \ \ \ \bar\nu_\ell N\ra\ell^++hadrons
\end{eqnarray}

\begin{figure}
\begin{minipage}{7.5cm}
\epsfig{file=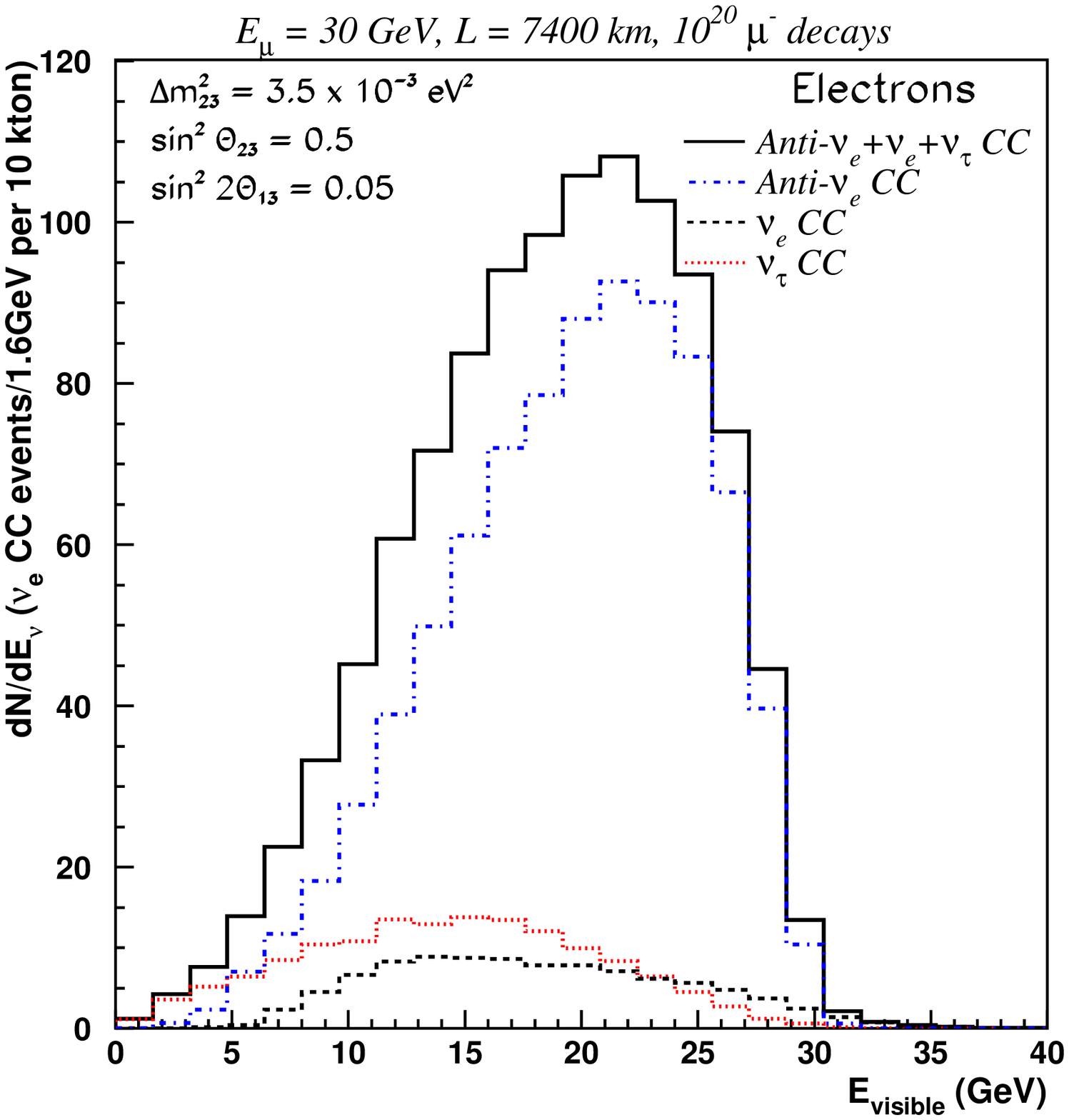,width=7.cm}
\caption{Visible energy spectrum for electron events:
$\nu_e$ CC (dashed line),  $\nu_\tau$ and
$\bar\nu_\tau$ (dotted line) and $\bar{\nu}_e$ CC (dot-dashed). 
The solid
histogram shows the sum of all contributions.}
\label{fig:elecont}
\end{minipage}
\begin{minipage}{7.5cm}
\epsfig{file=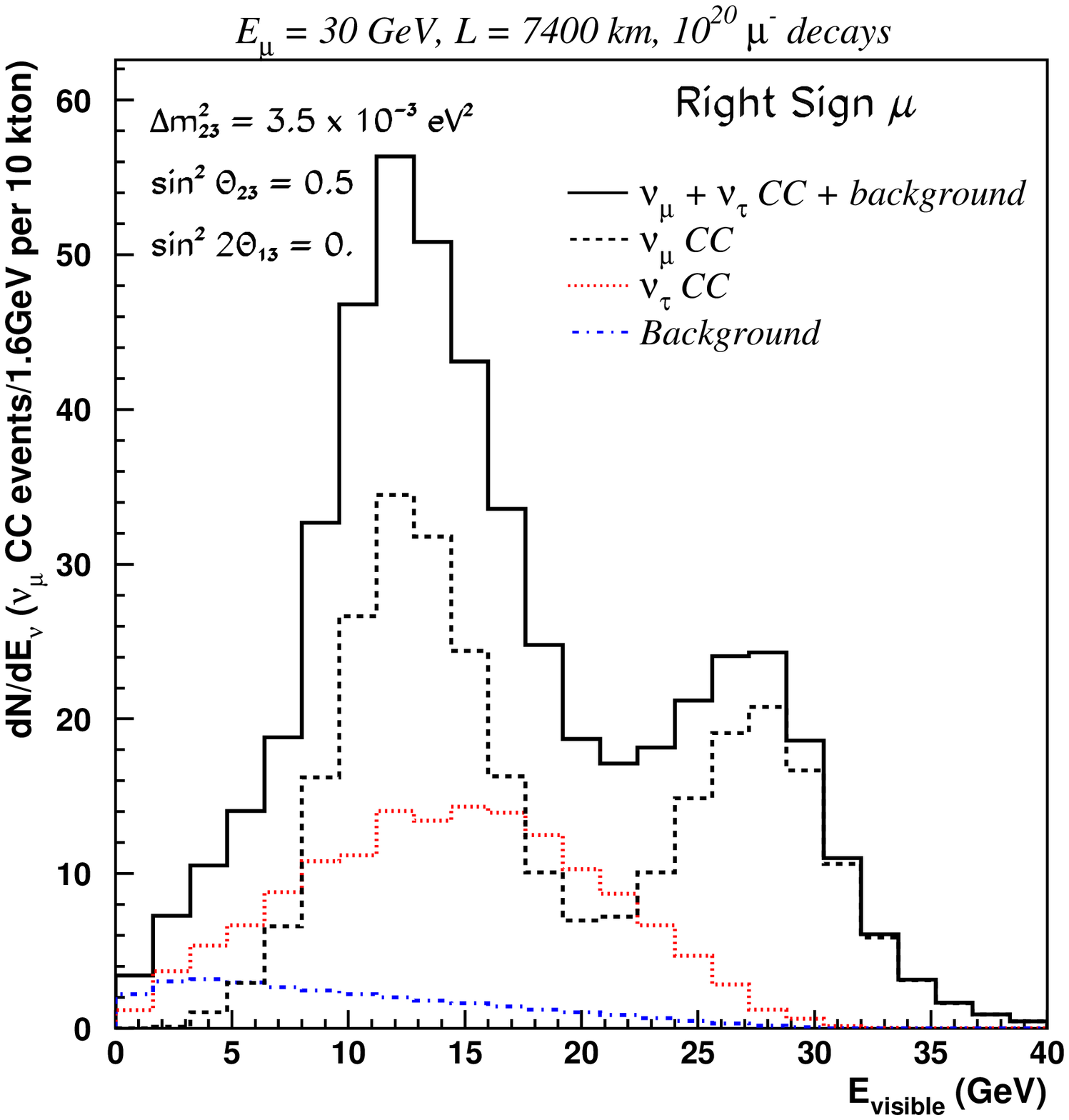,width=7.cm}
\caption{
same as Figure~\ref{fig:elecont} for right-sign muon
sample: $\nu_\mu$ CC (dashed line),  $\nu_\tau$ and
$\bar\nu_\tau$ (dotted line) and meson decay background (dot-dashed). 
The solid
histogram shows the sum of all contributions.}
\label{fig:dipevol}
\end{minipage}
\begin{minipage}{7.5cm}
\epsfig{file=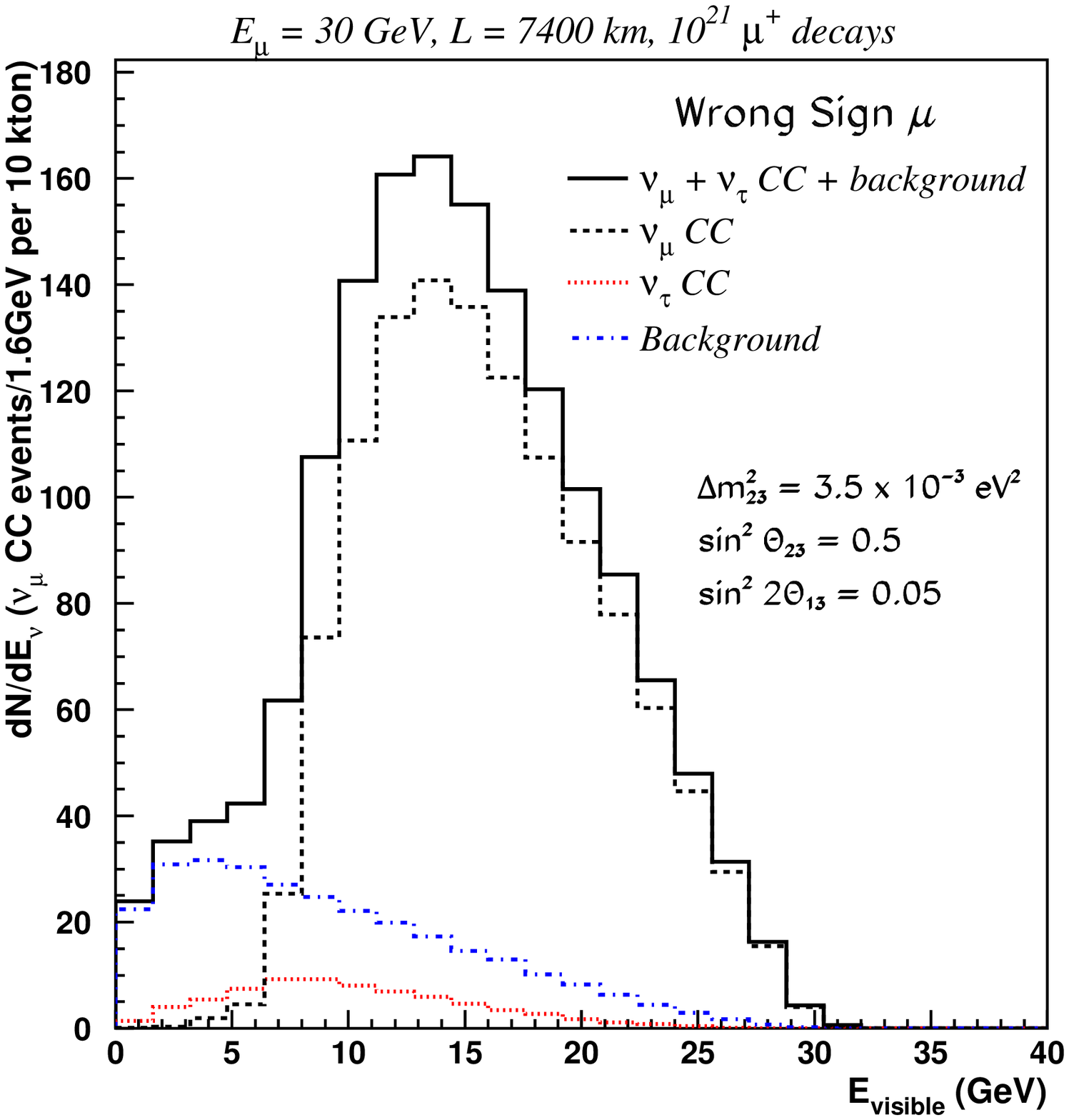,width=7.cm}
\caption{
same as Figure~\ref{fig:dipevol} for wrong sign muon
sample: $\nu_\mu$ CC (dashed line),  $\nu_\tau$ and
$\bar\nu_\tau$ (dotted line) and meson decay background (dot-dashed). 
The solid
histogram shows the sum of all contributions.}
\label{fig:wrongmu}
\end{minipage}
\begin{minipage}{7.5cm}
\epsfig{file=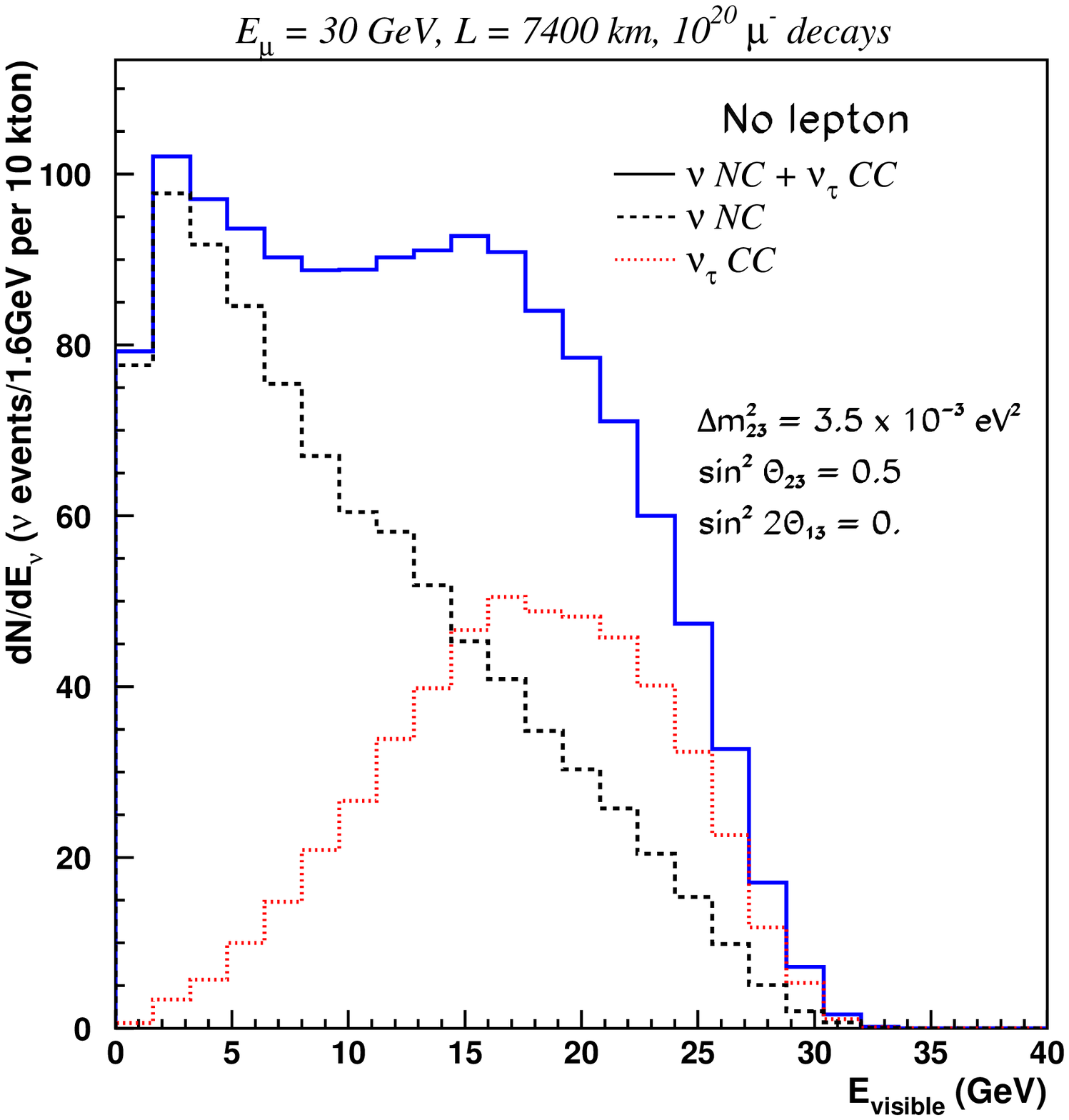,width=7.cm}
\caption{
same as Figure~\ref{fig:dipevol} for the no-lepton
sample: $\nu$ NC (dashed line),  $\nu_\tau$ and
$\bar\nu_\tau$ (dotted line).
The solid
histogram shows the sum of all contributions.}
\label{fig:nccont}
\end{minipage}
\end{figure}

We illustrate this in the case of a non-magnetized ICARUS-like detector.
Figures~\ref{fig:elecont},~\ref{fig:dipevol},~\ref{fig:wrongmu} 
and~\ref{fig:nccont} show the reconstructed visible energy
at the baseline $L=7400$km normalized to $10^{20}\mu$'s
for each event class for a specific oscillation scenario
with $\Delta m_{32}^2 = 3.5 \times 10^{-3}$ eV$^2$, 
$\sin^2 \tet23 = 0.5$ and $\sin^2 2\theta_{13} = 0.05$.
The different contributions including backgrounds for each 
event class have been evidenced in the plots. 
For example, in Figure~\ref{fig:dipevol}, the different processes that
contribute to the right-sign muon class are
unoscillated muons, taus and background events. 

Hence, the {\it ideal detector at the neutrino factory should possess the
following characteristics}:
\begin{itemize}
\item {\bf Particle identification}: the detector should be able to identify and
measure the leading charged lepton of the interaction, in order to tag the
incoming neutrino flavor.
\item {\bf Charge identification}: the sign of the leading lepton charge should
be measured, since it tags the helicity of the incoming neutrino.
\item {\bf Energy resolution}: the incoming neutrino energy $E_\nu$ is
reconstructed as $E_\nu = E_\ell+E_{had}$, where $E_\ell$ is the leading
lepton energy and $E_{had}$ is the hadronic energy. Hence, detector with
better energy resolution will reconstruct the parameter of the incoming
neutrino better, and therefore the oscillation probability.
\item {\bf Low energy threshold}: the reconstruction and identification
should be fully efficient for neutrino events below 15~GeV, as it is in
this region that we expect the cleanest and most ambiguous signal from $CP$
and $T$ violation, as we will demonstrate in the Sections 9 to 12.
\item {\bf Isotropic}: one might want to perform various similar experiments at different
baselines. The probably most efficient way to achieve this is to build a
large neutrino detector, {\it isotropic in nature}, capable of measuring
equally well
neutrinos from different sources located at different baselines
$L$. Because of the spherical shape of the Earth, sources located at different
baselines $L$ will reach the detector ``from below'' at different angles.
\end{itemize}

It should be stressed that these features need to be implemented on
detectors which have to be very massive. Indeed, the neutrino factory
will clearly be more intense (by at least one order of magnitude) than
current neutrino beams. However, the requirement to make precision
measurements and the considered rather long baselines (neutrino flux scales
as $L^{-2}$), 
implies that detector
in the range of 10~kt or more will be required.

\section{Detection of muons, electrons, taus and neutral currents}
Clearly, the goal to identify and measure all possible final state leptons
produced in charged current neutrino interactions as well as neutral
current neutrino interactions imposes some constraints on the detector
technology.

The most stringent problem is related to the achievable granularity in a
given detector configuration. Indeed, the requirement that detectors should
have a fiducial mass beyond the kton range immediately brings in various
choices and optimizations, since in first approximation, the total cost of
the detector will depend on its mass times its granularity. It is therefore
possible to opt for a poor granularity detector of very large mass, or a
high granularity detector of a smaller mass. In between these two extremes
one can have a continuous set of possible optimizations.

When we consider the possibility to tag the outgoing lepton, we can
subdivide the requirements as a function of the lepton type:
\begin{itemize}
\item Muons: the detection of muons is straight-forward. In principle, one looks for
penetrating particles. In practice, one has however to be careful of
backgrounds (misidentification) coming from $\pi^\pm$, $K^\pm$ decays and
from charm semileptonic decays.
\item Electrons: the detection of electrons is harder. In principle, one looks for large
and ``short'' energy deposition. In practice, one needs to carefully
separate electrons from $\pi^0$ conversions. Different levels of expertise
have been developed in the field, yielding typically background rejections
ranging from 1-2\% (for the worst granularity) to better than $10^{-3}$ for
the best granularity.
\item Taus: the detection of tau leptons is the hardest. One can attempt to
look on an-event-by-event basis for the tau ``kink''; this however requires an extremely good
reconstruction of the neutrino vertex. The required resolution has so far
only been achieved with the help of photographic emulsions, which pose
stringent constraints in scanning, and are strongly cost limited.
Another possibility is to provide a ``statistical separation''. In
particular, 60\% of the tau decays are into 1-prong, 3-prong or more
hadrons and hence look like neutral current events (i.e. without final
state leading electron or muon).
\end{itemize}
As far as the detection of the lepton charge is concerned, here also the
level of difficulty depends strongly on the type of lepton. In general, one
has to rely on a magnetic analysis to measure the charge.
\begin{itemize}
\item Muons: the measurement of the muon charge is relatively easy, since
tracks produced by muons are long. One can either envisage a fully
magnetized target or rely on a down-stream spectrometer.
\item Electrons: the measurement of the electron charge is the hardest. One
needs to measure significantly precisely the bending in the magnetic field
before the start of the electromagnetic shower. Hence, the measurement is
typically limited to a few $X_0$.
\item Taus: when the tau decays into a muon, then the measurement is
easy. Otherwise, one needs to rely on a very high granularity and
magnetized target, in order to identify the tau, its decay products and
eventually reconstruct the mother charge from the charge of the decay
products.
\end{itemize}

\section{Measurement of the electron charge in liquid argon}
\label{sec:elechar}
We saw that liquid argon imaging provides very good
tracking with $dE/dx$ measurement, and excellent calorimetric 
performance for contained showers. This allows for a very
precise determination of the energy of the particles in
an event. This is particularly true for electron
showers, which energy is very precisely measured. 

The possibility to complement these features with those
provided by a magnetic field has been considered. Embedding
the volume of argon into a magnetic field would not
alter the imaging properties of the detector and
the measurement of the bending of
charged hadrons or penetrating
muons would allow a
precise determination of the momentum and 
a determination of their charge.

\begin{figure}[htb]
\centering
\epsfig{file=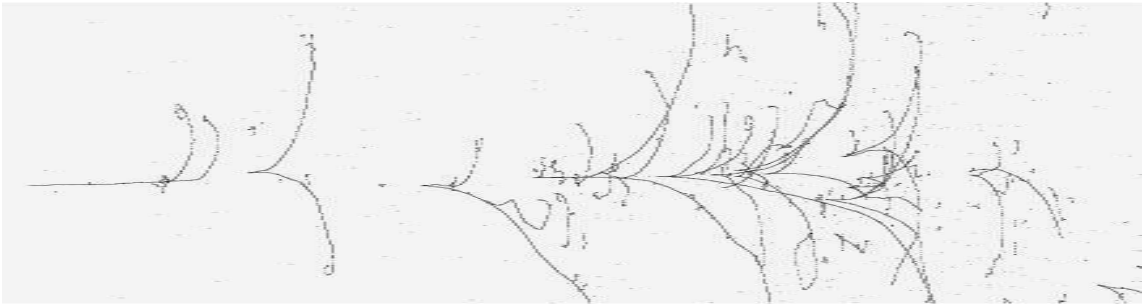,width=15cm,height=4cm}
\caption{Magnetized liquid argon TPC: simulation of the 2.5 GeV electron
shower in liquid argon. The field has a strength B=1.5~T and is directed
perpendicular to the sheet-plane.}
\label{eshower}
\end{figure}

We have recently started studying the effect of the magnetic field
on electrons (see Figure~\ref{eshower}). 
Unlike muons or hadrons, the early showering of electrons 
makes their charge identification difficult.
We however found that the determination
of the charge of electrons of energy in the range between
1 and 5 GeV is feasible with good
purity, provided the field has a strength in the range of 1~Tesla.
Preliminary estimates show that
these electrons exhibit an average curvature 
sufficient to have electron charge discrimination better than
$1\%$ with an efficiency of 20\%.


\section{CP and T violation measurements}

In a three-family neutrino oscillation scenario,
the flavor eigenstates
$\nu_\alpha(\alpha= e,\mu,\tau)$ are related to the mass eigenstates
$\nu'_i(i=1,2,3)$ by the mixing matrix U
\begin{equation}
\nu_\alpha=U_{\alpha i}\nu'_i
\end{equation}
and it is customary to parameterize it as:
\begin{equation}
U(\theta_{12},\theta_{13},\theta_{23},\delta)=\left(
\begin{tabular}{ccc}
$c_{12}c_{13}$      & $s_{12}c_{13}$   &  $s_{13}e^{-i\delta}$ \\
$-s_{12}c_{23}-c_{12}s_{13}s_{23}e^{i\delta}$ &
$c_{12}c_{23}-s_{12}s_{13}s_{23}e^{i\delta}$ & $c_{13}s_{23}$ \\
$s_{12}s_{23}-c_{12}s_{13}c_{23}e^{i\delta}$ &
$-c_{12}s_{23}-s_{12}s_{13}c_{23}e^{i\delta}$ & $c_{13}c_{23}$ 
\end{tabular}\right)
\end{equation}
with $s_{ij}=\sin\theta_{ij}$ and $c_{ij}=\cos\theta_{ij}$. 

We concentrate on transitions between electron
and muon neutrinos. The oscillation probability is
\begin{eqnarray}
 P(\nu_e\ra\nu_\mu)   = P(\bar\nu_\mu\ra\bar\nu_e)   = \nonumber \\ 
4c^2_{13}
\Bigl[  \sin^2\Delta_{23} s^2_{12} s^2_{13}
      s^2_{23} + c^2_{12}
      \left( \sin^2\Delta_{13}s^2_{13}s^2_{23} + 
        \sin^2\Delta_{12}s^2_{12}
         \left( 1 - \left( 1 + s^2_{13} \right) s^2_{23} \right)  \right)  \Bigr]  
\nonumber \\ 
  -  \frac{1}{2}c^2_{13}\sin (2\theta_{12})s_{13}\sin (2\theta_{23}) \cos\delta 
     \left[ \cos 2\Delta_{13} - \cos 2\Delta_{23} - 
       2\cos(2\theta_{12})\sin^2\Delta_{12}\right]  \nonumber \\ 
  + \frac{1}{2}c^2_{13}\sin\delta 
\sin(2\theta_{12})s_{13}\sin (2\theta_{23})
\left[ \sin2\Delta_{12} - \sin2\Delta_{13} + 
\sin 2\Delta_{23} \right]
\end{eqnarray}
where $\Delta_{jk}\equiv
\Delta m^2_{jk}L/4E_\nu$ (in natural units).
This expression has been split in a first part independent from the phase
$\delta$, and in the two parts proportional respectively to $\cos\delta$ and $\sin\delta$.
To obtain the probabilities for $\numu\ra\nue$ and $\bar\nue\ra\bar\numu$,
we must replace $\delta\longrightarrow -\delta$, with the effect of
changing $\sin\delta\longrightarrow -\sin\delta$ and 
$\cos\delta\longrightarrow \cos\delta$. {\it The term proportional to
$\sin\delta$ is the CP- or T-violating term, while 
the $\cos\delta$ term equally modifies the probability for 
both $CP$-conjugate states.}

From this dependence, we see that a precise measurement of
the $\nue\ra\numu$ oscillation probability can yield information
of the $\delta$-phase provided that the other oscillation
parameters in the expression are known sufficiently accurately.

The dependence of the parameter $\delta$ is a priori most
``visible'' in the energy-baseline range such that 
$|\Delta_{12}|=|\Delta m^2_{21}|L/4E_\nu\simeq 1$ and
$|\Delta_{23}|=|\Delta m^2_{23}|L/4E_\nu\simeq 1$. 

When $|\Delta_{12}|\ll 1$ and $|\Delta_{23}|\simeq 1$, we obtain
\begin{eqnarray}
 P(\nu_e\ra\nu_\mu) & \simeq & \frac{1}{2}c^2_{13} \Bigl\{ c^2_{13}\Delta^2_{12}+
 2s^2_{13}\left(\sin^2 \Delta_{13}+\sin^2\Delta_{23} \right) \\ \nonumber
& &  +2\Delta_{12}s_{13}\Bigl[ 
\sin(\Delta_{13}+\Delta_{23})\cos\delta+\left(1-\cos(\Delta_{13}+\Delta_{23})\right)\sin\delta 
\Bigr]
\Bigr\}
\end{eqnarray}

At even higher $E_\nu$ or smaller $L$, we further have, when 
both $|\Delta_{12}|\ll 1$ and $|\Delta_{13}|,|\Delta_{23}|\ll 1$:
\begin{eqnarray}
 P(\nu_e\ra\nu_\mu) & \simeq & \frac{1}{2}c^2_{13} \Bigl\{ c^2_{13}\Delta^2_{12}+
 2s^2_{13}\left(\Delta^2_{13}+\Delta^2_{23} \right) 
 +2\Delta_{12}s_{13}
(\Delta_{13}+\Delta_{23})\cos\delta
\Bigr\}
\label{eq:probhighe}
\end{eqnarray}
and the dependence on the phase is only through $\cos\delta$. From this
follows a degeneracy under the change of sign of $\delta$.
In this $L$ and $E_\nu$ range, a precise
determination of the oscillation probability can no longer determine the
sign of $\delta$.

\begin{figure}[tb]
\centering
\epsfig{file=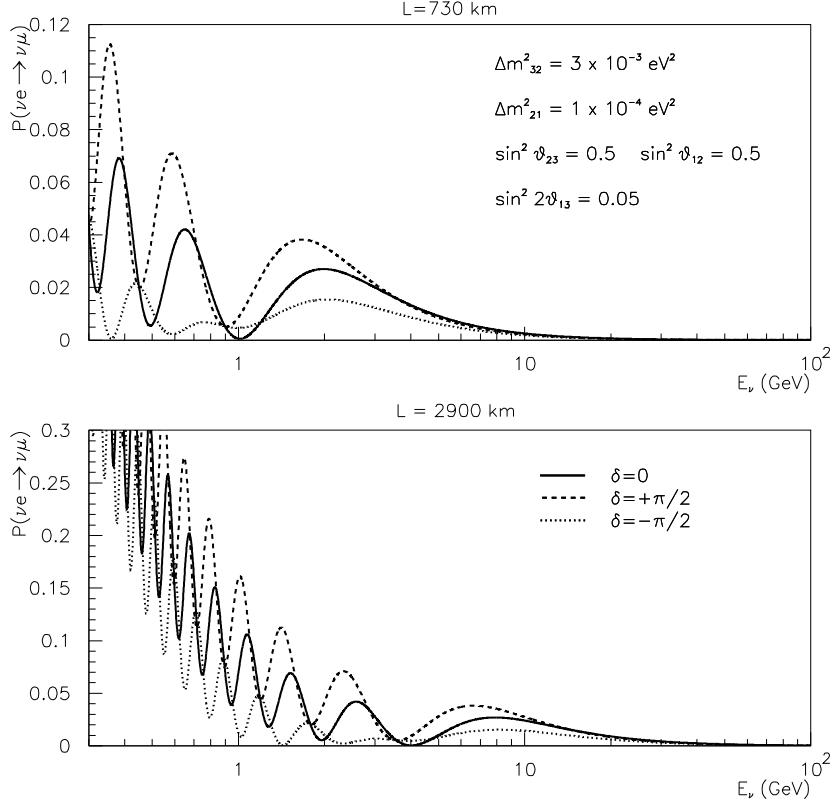,width=12cm}
\caption{Probability for $\nue\ra\numu$ oscillations in vacuum for two
baselines $L=730\ \rm km$ and 2900~km as a function of neutrino energy
$E_\nu$.
The probabilities are computed for three values of the $\delta$-phase:
$\delta=0$ (line), $\delta=+\pi/2$ (dashed), $\delta=-\pi/2$ (dotted).
The other oscillation parameters are $\Delta m^2_{32}=3\times 10^{-3}\ \rm
eV^2$, $\Delta m^2_{21}=1\times 10^{-4}\ \rm
eV^2$, $\sin^2 \theta_{23} = 0.5$, $\sin^2 \theta_{12} = 0.5$, 
and $\sin^2 2\theta_{13} = 0.05$.}
\label{fig:pnenmvac}
\end{figure}

The behavior at various energies and baselines is explicitly shown in
Figure~\ref{fig:pnenmvac} for the two baselines $L=730\ \rm km$ and
2900~km as a function of neutrino energy
$E_\nu$.
The probabilities are computed for three values of the $\delta$-phase:
$\delta=0$ (line), $\delta=+\pi/2$ (dashed), $\delta=-\pi/2$ (dotted).
The other oscillation parameters are: $\Delta m^2_{32}=3\times 10^{-3}\ \rm
eV^2$, $\Delta m^2_{21}=1\times 10^{-4}\ \rm
eV^2$, $\sin^2 \theta_{23} = 0.5$, $\sin^2 \theta_{12} = 0.5$, 
and $\sin^2 2\theta_{13} = 0.05$.

A region corresponding to the oscillation of the ``first maximum'' is clearly
visible on the curves. We define the energy  of the ``first maximum'' as
follows
\begin{eqnarray}
\label{eq:oscmax}
\Delta_{32}=\frac{\pi}{2} & \longrightarrow &
E^{max}_\nu\equiv \frac{\Delta m^2_{32}}{2\pi}L \nonumber \\
& \longrightarrow & E^{max}_\nu(\rm GeV)\simeq\Delta m^2_{32}(\rm
eV^2)\left(\frac{2\times 1.27}{\pi}\right){L(\rm km)}
\end{eqnarray}
which yields $E^{max}_{\nu}\simeq 2\rm\ GeV$ at 730~km,
$E^{max}_{\nu}\simeq 8\rm\ GeV$ at 2900~km and
$E^{max}_{\nu}\simeq 20\rm\ GeV$ at 7400~km
for $\Delta m^2_{32}=3\times 10^{-3}\rm\ eV^2$. This energy corresponds to
the point of maximum oscillation induced by $\Delta m^2_{32}$ and coincides
with the maximum when $\delta=0$. It will be useful when we discuss the
point of maximum sensitivity to the $\delta$-phase.

\begin{figure}[tb]
\centering
\epsfig{file=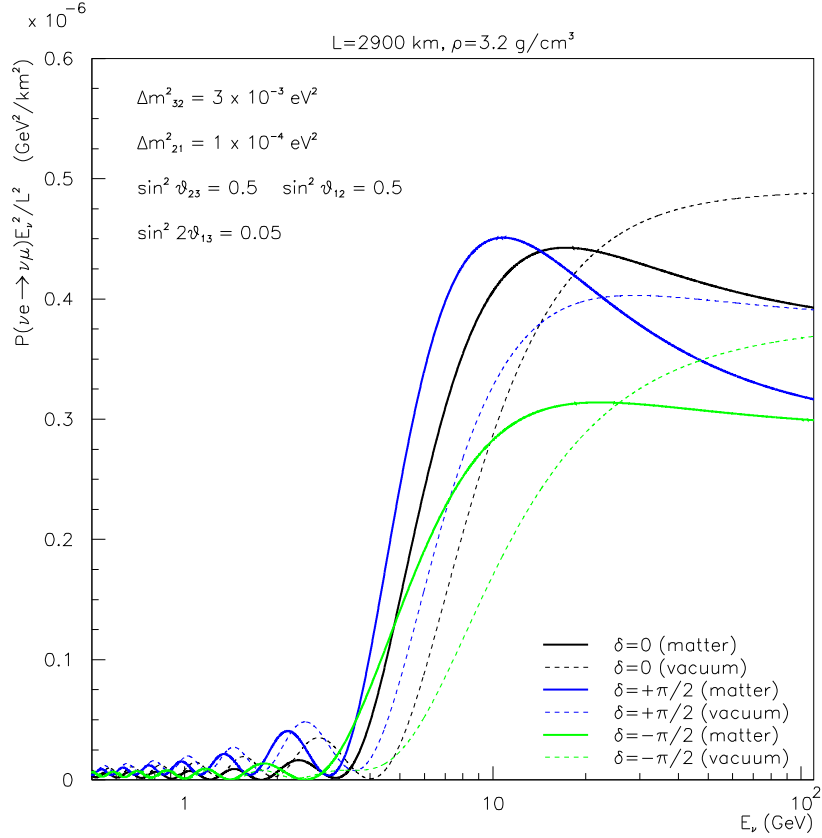,width=12cm}
\caption{Rescaled probability (see text) for $\nue\ra\numu$ oscillations for a
baseline $L=2900\ \rm km$ as a function of neutrino energy
$E_\nu$.
The probabilities are computed for neutrinos in matter (full line)
and in vacuum (dotted line), and for three values of the 
$\delta$-phase: $\delta=0$, $\delta=+\pi/2$, $\delta=-\pi/2$.
The other oscillation parameters are $\Delta m^2_{32}=3\times 10^{-3}\ \rm
eV^2$, $\Delta m^2_{21}=1\times 10^{-4}\ \rm
eV^2$, $\sin^2 \theta_{23} = 0.5$, $\sin^2 \theta_{12} = 0.5$, 
$\sin^2 2\theta_{13} = 0.05$.}
\label{fig:respcp1}
\end{figure}

\section{The rescaled probabilities}
In order to compare effects at different energies and various baselines, we
define a ``rescaled probability'' parameter that allows a direct comparison 
of effects.
Since to a good approximation, the neutrino 
energy distribution at the neutrino factory behaves like $E_\nu^2$ and
in addition, the neutrino flux scales like $L^{-2}$ due to the beam divergence,
we define the ``rescaled probability'' parameter $p(\nu_\alpha\osc\nu_\beta;E_\nu,L)$ as
\begin{equation}
p(\nu_\alpha\osc\nu_\beta;E_\nu,L)\equiv P(\nu_\alpha\osc\nu_\beta;E_\nu,L)\times \frac{E^2_\nu}{L^2}
\end{equation}
This behaviour is illustrated in Figure~\ref{fig:respcp1} for a baseline
$L=2900\rm\ km$.
We note that
(1) it approximately correctly ``weighs'' the probability by the neutrino energy
spectrum $E_\nu^2$ of the neutrino factory spectrum;
(2) it can be directly compared at different baselines, since it contains
the $L^{-2}$ attenuation of the neutrino flux with distance $L$;
(3) $p$ tends to a constant for $E_\nu\rightarrow \inf$, hence
the high energy behavior can be easily studied.

\section{Propagation in matter}
Since neutrino factories will be associated to long baseline, it is not
possible to avoid 
including effects associated to the neutrino propagation through the Earth matter.
The simplest way to take into account these effects is to maintain the
formalism developed for propagation in vacuum and to replace the mixing angles and the neutrino
mass differences by ``effective'' values. 

An important quantity for
matter effects is $D$, defined as
\begin{equation}
D(E_\nu,\rho)\equiv 2\sqrt{2}G_F n_e E_\nu=7.56\times 10^{-5}eV^2(\frac{\rho}{g cm^{-3}})
(\frac{E_\nu}{GeV})
\end{equation}
where $n_e$ is the electron density and $\rho$ the matter density. For 
antineutrinos, $D$ is replaced by $-D$. 

There are two specific
neutrino energies of interest when neutrinos propagate through matter:
\begin{enumerate}
\item for $D\approx \Delta m^2_{32}$, we reach for neutrinos the MSW
resonance, in which the effective mixing angle
$\sin^2(2\theta^m_{13})\approx 1$. In terms of neutrino energy, this
implies
\begin{eqnarray}
E^{res}_\nu& = &\frac{\cos2\theta_{13}\Delta m^2_{32}}{2\sqrt{2}G_F
n_e} = {\cal E}^{res}\cos2\theta_{13}\Delta m^2_{32} \nonumber \\
& \simeq &\frac{1.32\times 10^4\cos2\theta_{13}\Delta
m^2_{32}(\rm eV^2)}{\rho(g/cm^3)}\ \rm in\ GeV
\end{eqnarray}
where ${\cal E}^{res}=(2\sqrt{2}G_Fn_e)^{-1}$. For density parameters
$\rho$
equal to $2.7$, $3.2$ and $3.7\rm\ g/cm^3$ one finds
$E^{res}_\nu \simeq 14.1$, $12.3$ and $10.7\rm\ GeV$ for $\Delta
m^2_{32}=3\times 10^{-3}\rm\ eV^2$.
\item for $D> 2 \Delta m^2_{32}$, the effective mixing angle for neutrinos
is always smaller than that in vacuum, i.e.
$\sin^2(2\theta^m_{13})< \sin^2(2\theta_{13})$.
In terms of neutrino energy, this
is equivalent to
\begin{equation}
E_\nu > 2E^{res}_\nu
\end{equation}
\item these arguments are independent of the baseline $L$ and depend only
on the matter density $\rho$.
\end{enumerate}

\section{Detecting the $\delta$ phase at the NF}
In order to further 
study the dependence of the $\delta$-phase, we consider the
following three quantities which are good discriminators for a
non-vanishing 
phase $\delta$:
\begin{enumerate}
\item $\Delta_\delta \equiv  P(\nu_e\ra\nu_\mu,\delta=+\pi/2)-
P(\nu_e\ra\nu_\mu,\delta=0)$\\
The discriminant $\Delta_\delta$ can be used in an experiment where
one is comparing the measured $\nue\ra\numu$ oscillation probability as a
function of the neutrino energy $E_\nu$ compared to a ``Monte-Carlo
prediction'' of the spectrum in absence of $\delta$-phase.
\item $\Delta_{CP}(\delta) \equiv  P(\nu_e\ra\nu_\mu,\delta)-P(\bar\nu_e\ra\bar\nu_\mu,\delta)$ \\
The discriminant $\Delta_{CP}$ can be used in an experiment by
comparing the appearance of $\nu_\mu$ (resp. $\bar\nu_\mu$) in a beam of
stored $\mu^+$ (resp. $\mu^-$) decays as a function of the
neutrino energy $E_\nu$.
\item $\Delta_{T}(\delta) \equiv
P(\nu_e\ra\nu_\mu,\delta)-P(\nu_\mu\ra\nu_e,\delta)$ or
$\bar\Delta_{T}(\delta) \equiv  P(\bar\nu_e\ra\bar\nu_\mu,\delta)-
P(\bar\nu_\mu\ra\bar\nu_e,\delta)$ \\
The discriminant $\Delta_{T}$ can be used in an experiment by
comparing the appearance of $\nu_\mu$ (resp. $\bar\nu_\mu$) {\bf and}
$\bar\nu_e$ (resp. $\nu_e$) and in a beam of
stored $\mu^+$ (resp. $\mu^-$) decays as a function of the
neutrino energy $E_\nu$.
\end{enumerate}
Each of these discriminants have their advantages and disadvantages.

The $\Delta \delta$-method consists in searching for distortions in the
visible energy spectrum of events produced by the $\delta$-phase. While
this method can in principle provide excellent determination of the phase limited
only by the statistics of accumulated events, in practice, systematic
effects will have to be carefully kept under control in order to look for a
small effect in a seen-data versus Monte-Carlo-expected comparison.
In addition, the precise knowledge of the other oscillation parameters will
be important, and as will be discussed below, there is a risk of degeneracy
between solutions and a possible strong correlation with the $\theta_{13}$
angle at high energy.

The $\Delta_{CP}$ is quite straight-forward, since it involves comparing
the appearance of so-called wrong-sign muons for two polarities of the
stored muon beam. 
It really takes advantage from the fact that
experimentally energetic muons are rather easy to detect and identify due
to their penetrating nature, and with the help of a magnetic field, their
change can be easily measured, in order to suppress the non-oscillated
background from the beam. 
A non-vanishing $\Delta_{CP}$ should in principle be a direct proof
for a non-vanishing $\delta$-phase.
This method suffers, however, from the inability
to perform long-baseline experiment through vacuum. Indeed, matter effects
will largely ``spoil'' $\Delta_{CP}$ since it involves both neutrinos and
antineutrinos, which will oscillate very differently
through matter. Hence, the $\Delta_{CP}$ requires a good understanding of
the effects related to matter. In addition, it involves measuring neutrinos
and antineutrinos. The matter suppression of the antineutrinos will in
practice determine the statistical accuracy with which the discriminant can
be measured.

Finally, the $\Delta_{T}$ is the theoretically cleanest method, since it
does not suffer from the problems of $\Delta \delta$ and
$\Delta_{CP}$. Indeed, a difference in oscillation probabilities
between $\nue\ra\numu$ and
$\numu\ra\nue$ would be a direct proof
for a non-vanishing $\delta$-phase. In addition, matter affects both
probabilities in a same way, since it involves only neutrinos.
Unfortunately, it is experimentally very challenging to discriminate the
electron charge produced in the events, needed in order to suppress the
background from the beam. However, one can decide to measure only
neutrinos, which are enhanced by matter effects, as opposed to
antineutrinos in the $\Delta_{CP}$ which were matter suppressed, and hence
the statistical accuracy of the measurement will be determined by the
efficiency to recognize the electron charge, rather by matter suppression.

\section{The $L/E_\nu$ scaling of the CP and T effects}
Regardless of their advantages and disadvantages, there is one thing in
common between the three discriminants $\Delta_\delta$, 
$\Delta_{CP}$ and $\Delta_{T}$: their behavior with respect to the
neutrino energy $E_\nu$ and the baseline $L$.
By explicit calculation, we find
\begin{eqnarray}
\label{eq:deldeldef}
\Delta_\delta&=&-\frac{1}{2}c^2_{13}\sin 2\theta_{12}s_{13}\sin
2\theta_{23}\times
\\ \nonumber
& & \Bigl[
\cos 2\Delta_{13}-\cos2\Delta_{23}-2\cos2\theta_{12}\sin^2\Delta_{12}
+\sin 2\Delta_{12}-\sin 2\Delta_{13}+\sin 2\Delta_{23}\Bigr] \\ \nonumber
&=& -\frac{1}{2}c^2_{13}s_{13}
\Bigl[
\cos 2\Delta_{13}-\cos2\Delta_{23}
+\sin 2\Delta_{12}-\sin 2\Delta_{13}+\sin 2\Delta_{23}\Bigr] 
\end{eqnarray}
where for the second line
we assumed for simplicity $\theta_{12}=\theta_{23}=\pi/4$, and similarly,
\begin{eqnarray}
\label{eq:delcptdef}
\Delta_{CP}=\Delta_{T} & = &
c^2_{13}s_{13}\sin2\theta_{12}\sin 2\theta_{23}\sin\delta
\Bigl[ \sin 2\Delta_{12}- \sin 2\Delta_{13} + \sin 2\Delta_{23}\Bigr]
\nonumber \\
& = &
-c^2_{13}s_{13}\sin2\theta_{12}\sin 2\theta_{23}\sin\delta
\Bigl[ \sin \Delta_{12}\sin \Delta_{13}\sin\Delta_{23}\Bigr]
\end{eqnarray}
As expected, both expressions vanish in the limit 
$\Delta m^2_{12}\ra 0$ where $\Delta m^2_{13}\ra \Delta m^2_{23}$.
Also, as one reaches the higher energies,
the terms $\Delta_{CP}=\Delta_{T}$ vanish as 
\begin{eqnarray}
\label{eq:delcptdefvac}
|\Delta_{CP}|=|\Delta_{T}| 
& \simeq & c^2_{13}s_{13}\sin2\theta_{12}\sin 2\theta_{23}\sin\delta
 \Delta m^2_{12}\left(\frac{L}{4E_\nu}\right)
\sin^2\left(\Delta m^2_{23}\frac{L}{4E_\nu}\right) \nonumber \\
& \simeq & c^2_{13}s_{13}\sin2\theta_{12}\sin 2\theta_{23}\sin\delta
\Delta m^2_{12}(\Delta m^2_{23})^2
\left(\frac{L}{4E_\nu}\right)^3
\end{eqnarray}
hence, in the very high energy limit at fixed baseline, the effects
decrease as $E_{\nu}^{-3}$. That the effects disappear at high energy is
expected, since in this regime, the ``oscillations'' of the various
$\Delta_{jk}$'s wash out.

{\it The important point is that all expressions depend from some factors which
contain the various mixing angles, multiplied by oscillatory terms which
always vary like sine or cosine of $\Delta_{jk}$-terms} (the terms in
squared brackets in the expressions above). 
{\it Hence, we expect the various discriminants to scale like
$\Delta_{jk}\propto L/E_\nu$. }
The sensitivity to the $\delta$-phase will therefore follow the
behavior of the oscillation probability, and we therefore argue that the
maximum of the effect will occur around the ``first maximum'' of the
oscillations, i.e. for
$E^{max}_\nu\equiv \Delta m^2_{32}L/2\pi$ (see Eq.~(\ref{eq:oscmax}))
\footnote{Strictly speaking, the maximum of the $\delta$-phase
sensitivity does not lie exactly at the ``first maximum'' as defined in 
Eq.~(\ref{eq:oscmax}). From Eq.~(\ref{eq:delcptdef}), we expect the
maximum to be ``shifted'' to higher values of $L/E_\nu$, since it
corresponds to the maximum of the term
\begin{equation}
 \sin \Delta_{12}\sin \Delta_{13}\sin\Delta_{23}\simeq 
 \Delta m^2_{12}\frac{L}{4E_\nu}\sin^2\left(\Delta m^2_{23}\frac{L}{4E_\nu}\right)
\end{equation}
which has the functional form $x\sin^2 x$ and, therefore, has its maximum
shifted to higher values of $x$ compared to $\sin^2 x$. This small shift is
smaller than the oscillation wavelength itself, and does not cause a
major problem, since experimentally we will always have sufficient energy
range to cover the full oscillation.}

These considerations are strictly true only for propagation in vacuum. When
neutrinos propagate through matter, matter effects will alter these
conclusions. We will however show that as long as the baseline is smaller
than some distance such that the corresponding ``first maximum''
$E^{max}_\nu$ lies below
the MSW resonance neutrino energy $E^{res}_{\nu}$, the considerations
related to the $L/E_\nu$ scaling are still largely valid. 

{\it In what way does the matter effect alter the ability to look for effects
related to the $\delta$-phase?} It is incorrect to believe that only the
$\Delta_{CP}$ discriminant will be affected by propagation through matter,
since it is the only one to a priori mix neutrinos and antineutrinos. In
reality, the ``dangerous'' effect of matter is to reduce the dependence of
the probability on the $\delta$-phase, and this for any kind of discriminant.

In matter, we would for example write the $\Delta_T$ discriminant
as
\begin{eqnarray}
|\Delta^m_{T}| & = &
(c^{m}_{13})^2s^m_{13}\sin2\theta^m_{12}\sin 2\theta^m_{23}\sin\delta^m
\Bigl[ \sin \Delta^m_{12}\sin \Delta^m_{13}\sin\Delta^m_{23}\Bigr]
\nonumber \\
& \simeq &
(c^{m}_{13})^2s^m_{13}\sin2\theta^m_{12}\sin 2\theta_{23}\sin\delta
\Bigl[ \sin \Delta^m_{12}\sin \Delta^m_{13}\sin\Delta^m_{23}\Bigr]
\end{eqnarray}
where because of our choice of $\Delta m^2_{jk}$'s,
we have $\theta^m_{23}\approx \theta_{23}$ and
$\delta^m\approx \delta$. 
This implies that the $\delta$-phase discriminants have a different
structure that the terms that define the probability of the oscillation
(i.e. the non $\delta$-phase dependent terms).
The discriminants are the products of sines and cosines of {\it all} mixing
angles and of the $\Delta_{jk}$'s (see Eqs.~(\ref{eq:deldeldef}) and
(\ref{eq:delcptdef})). Because of this structure, their property in matter
is different.

The behavior for neutrino energies above the MSW
resonance $E^{res}_\nu$ is determined by the fact that in this energy
regime, $\theta^m_{13}(E>E_\nu^{res})\rightarrow \pi/2$ and therefore
$c^{m}_{13}\ra 0$ and $s^m_{13}\ra 1$. 
More explicitly, one can show that
\begin{eqnarray}
(c^{m}_{13})^2s^m_{13} & \propto & E_\nu^{-2} \nonumber \\
\sin 2\theta^m_{23} & \simeq & \sin 2\theta_{23} = const. \nonumber \\
\sin 2\theta^m_{12} & \ra & const. \nonumber \\
\Delta M^2_{31} & \approx & \Delta M^2_{32} \propto E_\nu \nonumber \\
\Delta M^2_{21} & \approx & \Delta m^2_{32} = const
\end{eqnarray}
Therefore,
\begin{eqnarray}
|\Delta^m_{T}| & \propto & 
E_\nu^{-2} \sin\delta 
\Bigl[ \sin\left( \Delta m^2_{32}\frac{L}{4E_\nu}\right)
\sin^2 \left( \Delta M^2_{13}\frac{L}{4E_\nu}\right)\Bigr]
\propto 
E_\nu^{-3}
\end{eqnarray}
and one recovers a neutrino energy dependence identical to that in vacuum
(see Eq.~(\ref{eq:delcptdefvac})). Note also that the argument of the sine
function $\Delta M^2_{13}L/4E_\nu$ is {\it not} small (i.e. the
approximation $\sin x\simeq x$ is not valid). For our choice of oscillation
parameters, the mass difference is approximately equal to 
$\Delta M^2_{13}(\rm eV^2)\simeq 3\times 10^{-4}\times E_\nu(\rm GeV)$,
and hence the dependence on the baseline is 
\begin{eqnarray}
|\Delta^m_{T}| & \propto & 
E_\nu^{-2} \sin\delta 
\Bigl[ \sin\left( 1.27\Delta m^2_{32}\frac{L(\rm km)}{E_\nu(\rm GeV)}\right)
\sin^2 \left( 3.8\times 10^{-4} L(\rm km)\right)\Bigr]
\end{eqnarray}
Hence, the discriminant will first be enhanced and then be suppressed by
matter effects. The maximum is found when the sine squared function reaches
a maximum, or at approximately 4000~km under the assumption of high energy
neutrinos. 

As anticipated, these discussions say that if one wants to study
oscillations in the region of the ``first maximum'', one should not choose
a too large baseline $L$, otherwise, matter effects will suppress the
oscillation probability. This is even more so true, as it will be recalled
below, that the magnitude of the effects related to the $\delta$-phase are
suppressed more rapidly than the oscillation.

The simplest way to express the condition on the matter is to
require that the energy of the ``first maximum'' be smaller than the MSW
resonance energy:
\begin{eqnarray}
2\sqrt{2}G_Fn_eE^{max}_\nu\lesssim\Delta m^2_{32}\cos 2\theta_{13}
\end{eqnarray}
and, by inserting the definition of $E^{max}_\nu\equiv\Delta m^2_{32}L/2\pi$ we get
\begin{eqnarray}
\label{eq:lmaxmatter}
L_{max}& \lesssim& \frac{\pi\cos 2\theta_{13}}{\sqrt{2}G_Fn_e} \approx
\frac{\pi\cos 2\theta_{13}}{2\times 1.27\times 7.56\times 10^{-5}(\rm
eV^2)\rho(\rm g/cm^3)} \nonumber \\ & \approx &\frac{1.5\times 10^4(\rm km)}{\rho(\rm
g/cm^3)}
\approx 5000\rm\ km
\end{eqnarray}

\begin{figure}[tb]
\centering
\epsfig{file=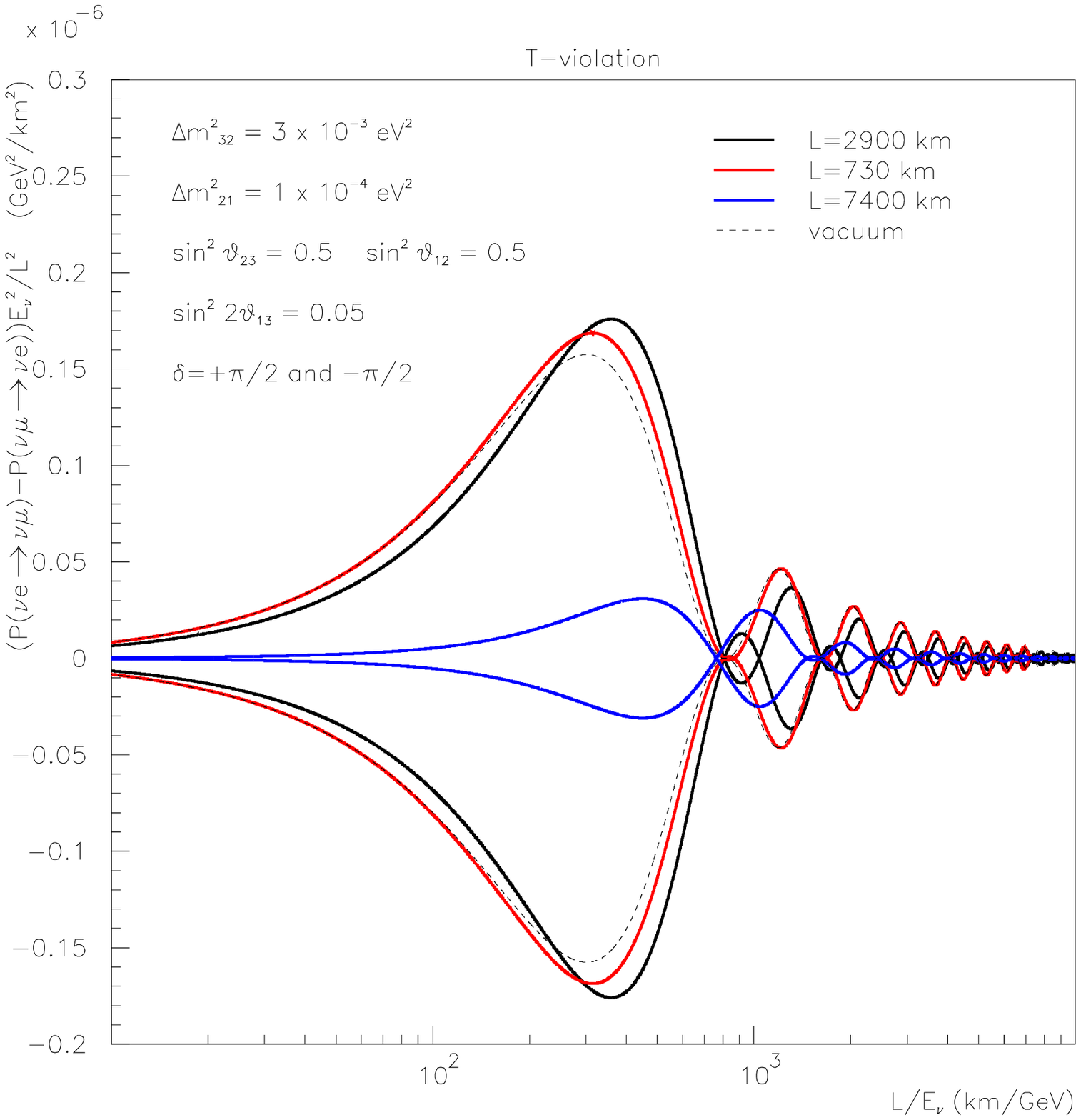,width=7cm}
\epsfig{file=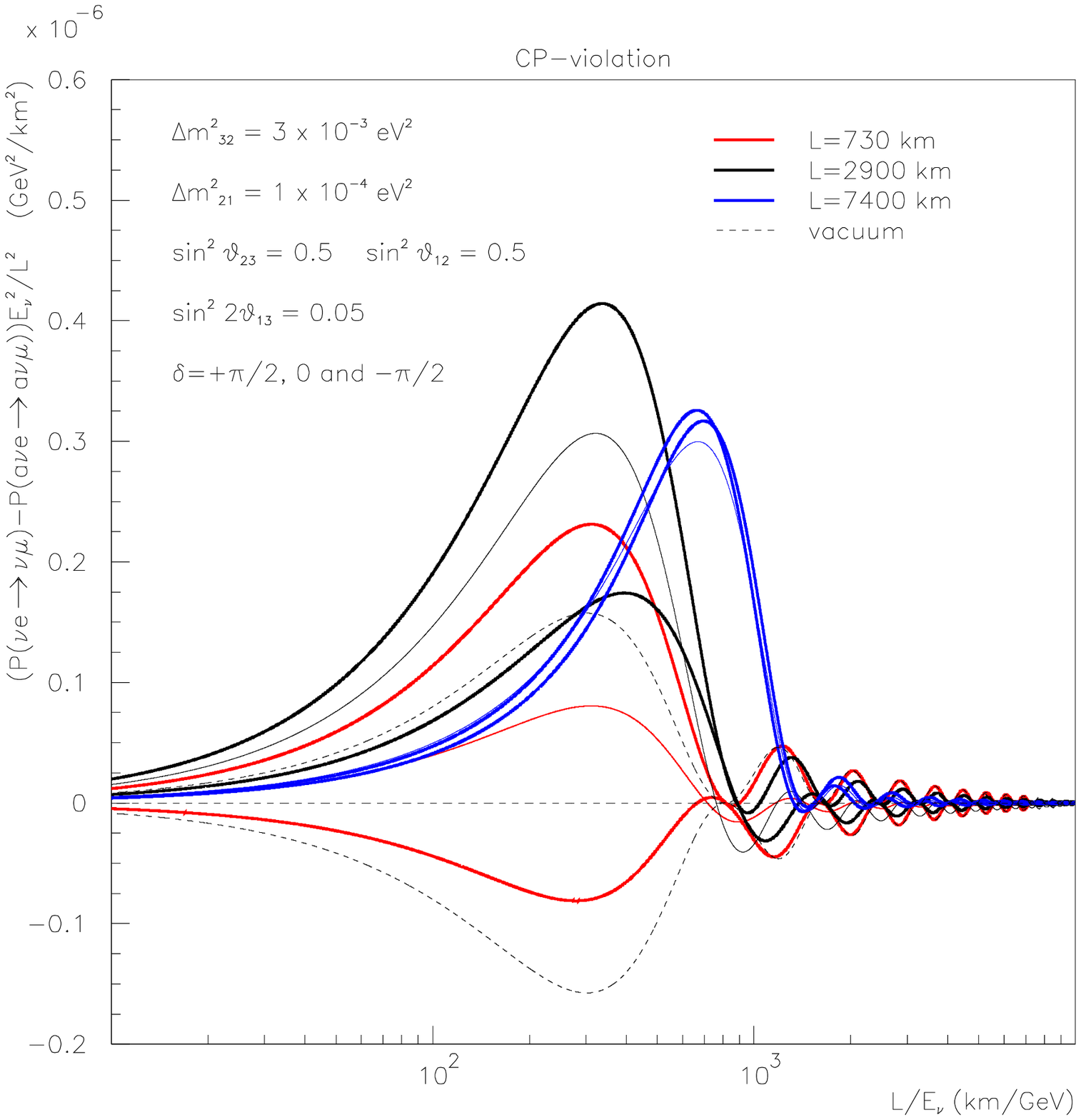,width=7cm}
\caption{The rescaled $\Delta_{T}$ and $\Delta_{CP}$ discriminants (see text for
definition) as a function of the $L/E_\nu$ ratio,
computed for neutrinos propagating in matter
at three different baselines $L=730\rm\ km$, 2900~km and 7400~km, and
also for propagation vacuum (independent of baseline).
Three sets of curves are represented, corresponding to $\delta=+\pi/2$ and
$\delta=-\pi/2$ (thick lines) and $\delta=0$ (thin lines).
The other oscillation parameters are $\Delta m^2_{32}=3\times 10^{-3}\ \rm
eV^2$, $\Delta m^2_{21}=1\times 10^{-4}\ \rm
eV^2$, $\sin^2 \theta_{23} = 0.5$, $\sin^2 \theta_{12} = 0.5$
and $\sin^2 2\theta_{13} = 0.05$.}
\label{fig:deltle}
\end{figure}

To summarize, {\bf we find that the discriminants of the $\delta$-phase all
scale with $L/E_\nu$}. 
This is illustrated in Figure~\ref{fig:deltle}, where
the rescaled $\Delta_{T}$ and $\Delta_{CP}$ discriminants are plotted
as a function of the $L/E_\nu$ ratio. In the plots, sets of curves are
shown for $\delta=+\pi/2$ and $\delta=-\pi/2$. We see that the rescaled
$T$-discriminant is (as expected) antisymmetric with respect to $\delta$. For the shorter
baselines (730~km and 2900~km) it is almost equivalent to the vacuum case
(dashed curves). The 7400~km baseline yields a highly suppressed $T$-effect.
The $CP$-discriminant has the same features, but is shifted with respect to
zero due to matter enhancement.

The most favorable choice of neutrino energy $E_\nu$ and
baseline $L$ is in the region of the ``first maximum'' given by 
$(L/E_\nu)^{max}\simeq 400$ for $|\Delta m^2_{32}|=3\times 10^{-3}\rm\
eV^2$. This leaves a great flexibility in the choice of the actual neutrino
energy and the baseline, since only their ratio $L/E_\nu$ is
determinant. 

{\it Because of the rising neutrino cross-section with energy,
it will be more favorable to go to higher energies if the
neutrino fluence is constant. Keeping the $L/E_\nu$ ratio constant, 
this implies an optimization at longer baselines $L$. One will hence gain
with the baseline $L$ until we reach $L_{max}\approx 5000\rm\ km$ beyond
which matter effects will effectively reduce the dependence of
the oscillation probabilities on the $\delta$-phase.}

\section{The correlation with $\theta_{13}$}
We begin the discussion with the detection of the $\delta$-phase with the
method of the $\Delta_\delta$ discriminant. We recall that this method
implies the comparison of 
the measured $\nue\ra\numu$ oscillation probability as a
function of the neutrino energy $E_\nu$ compared to a ``Monte-Carlo
prediction'' of the spectrum in absence of $\delta$-phase.

When searching for effects related to the $\delta$-phase by comparing
the measured $\nue\ra\numu$ oscillation probability as a
function of the neutrino energy $E_\nu$ to a ``Monte-Carlo
prediction'' of the spectrum in absence of $\delta$-phase, requires
necessarily a precise knowledge of the other oscillation parameters
entering in the oscillation probability expression.

In particular, the knowledge of the angle $\theta_{13}$ could be quite
important. Indeed, the $\nue\ra\numu$ oscillation is primarily driven by
the $\theta_{13}$ angle and {\it only thanks to a different energy dependence of
the terms proportional to $\delta$ than to those independent of
$\delta$ can one hope to determine $\theta_{13}$ and $\delta$ at the same
time!}

This is however not true at high energy, when 
both $|\Delta_{12}|\ll 1$ and $|\Delta_{13}|,|\Delta_{23}|\ll 1$.
This can be explicitly shown for example
for simplicity in the limit of small $\theta_{13}$. The rescaled
probability is in this case a constant:
\begin{eqnarray}
 p(\nu_e\ra\nu_\mu) & \simeq & 
\frac{(\Delta m^2_{12})^2}{32} \Bigl\{ 1+ 
 2Ms^2_{13} +8Ns_{13}\cos\delta
\Bigr\}
\end{eqnarray}
where $M=((\Delta m^2_{13})^2+(\Delta m^2_{23})^2)/(\Delta m^2_{12})^2$ and
$N=(\Delta m^2_{13}+\Delta m^2_{23})/(\Delta m^2_{12})$. The absence of
``oscillations'' at high energy implies that a change of $\theta_{13}$ can
mimic a change of $\delta$.
In practice, this implies that the best energy range to look for effects
related to the $\delta$-phase is close the ``first oscillation maximum'',
i.e. $E^{max}_{\nu}\simeq 2\rm\ GeV$ at 730~km or
$E^{max}_{\nu}\simeq 8\rm\ GeV$ at 2900~km. This implies that the detector
should be able to reconstruct neutrino events at those energies with high
efficiency, and low background. 

\section{Two concrete examples at L=730~km and 2900~km}
In order to assess with concrete examples the use of the $\delta$-phase
discriminants, we consider the two baselines, with corresponding muon beam energy:
\begin{itemize}
\item $L$=732 km, $E_\mu=7.5\rm\ GeV$, $10^{21}$ muon decays
\item $L$=2900 km, $E_\mu=30\rm\ GeV$, $2.5\times 10^{20}$ muon decays
\end{itemize}
Both examples were chosen to have the same $L/E_\mu$. Because of the linear
rise of the neutrino cross-section with $E_\nu$, the factor 4 in muon
energy between the 732~km and 2900~km case, is ``compensated'' by
an increase of intensity by the same factor in favor of the shorter baseline.

The expected event rates are shown in Tables~\ref{tab:rates1} and \ref{tab:rates2}.

\begin{table}
\begin{center}
\begin{tabular}{|ll|c|c|}\hline
&&$E_\mu=7.5\rm\ GeV$& $E_\mu=30\rm\ GeV$ \\
&Process& $L=732\rm\ km$& $L=2900\rm\ km$\\ 
& & $10^{21}$ $\mu^-$ & $2.5\times 10^{20}$ $\mu^-$ \\
\hline
& $\nu_\mu$ CC&41690&36050\\
Non-oscillated& $\nu_\mu $ NC&10700&10300\\
rates& $\bar{\nu}_e$ CC&14520&13835\\
& $\bar{\nu}_e $ NC&4266&4975\\ \hline
Oscillated& $\bar{\nu}_e\osc\bar{\nu}_\mu$ CC&88&50\\
events ($\delta=\pi/2$)& $\nu_\mu\osc\nu_e $ CC&258&238\\\hline
Oscillated& $\bar{\nu}_e\osc\bar{\nu}_\mu$ CC&100&54\\
events ($\delta=0$)& $\nu_\mu\osc\nu_e $ CC&385&333\\\hline
Oscillated& $\bar{\nu}_e\osc\bar{\nu}_\mu$ CC&100&55\\
events ($\delta=-\pi/2$)& $\nu_\mu\osc\nu_e $ CC&376&330\\ \hline
\end{tabular}
\end{center}
\caption{Event rates for a 10 kton detector. 
The oscillation parameters are: $\Delta m_{32}^2 = 3 \times 10^{-3}\rm\ eV^2$, 
$\Delta m^2_{12}=1\times 10^{-4}\rm\ eV^2$,
$\sin^2 \theta_{23} = 0.5$, $\sin^2 \theta_{12}=0.5$
and $\sin^2 2\theta_{13} = 0.05$.}
\label{tab:rates1}
\end{table}

\begin{table}
\begin{center}
\begin{tabular}{|ll|c|c|}\hline
&&$E_\mu=7.5\rm\ GeV$& $E_\mu=30\rm\ GeV$ \\
&Process& $L=732\rm\ km$& $L=2900\rm\ km$\\ 
& & $10^{21}$ $\mu^+$ & $2.5\times 10^{20}$ $\mu^+$ \\
\hline
& $\bar{\nu}_\mu$ CC&16570&15962\\
Non-oscillated& $\bar{\nu}_\mu $ NC&5096&5600\\
rates& $\nu_e$ CC&37570&32100\\
& $\nu_e $ NC&9143&9175\\ \hline
Oscillated& $\nu_e\osc\nu_\mu$ CC&445&397\\
events ($\delta=\pi/2$)& $\bar{\nu}_\mu\osc\bar{\nu}_e $ CC&86&46\\ \hline
Oscillated& $\nu_e\osc\nu_\mu$ CC&438&387\\
events ($\delta=0$)& $\bar{\nu}_\mu\osc\bar{\nu}_e $ CC&86&45\\ \hline
Oscillated& $\nu_e\osc\nu_\mu$ CC&289&277\\
events ($\delta=-\pi/2$)& $\bar{\nu}_\mu\osc\bar{\nu}_e $ CC&77&42\\ \hline
\end{tabular}
\end{center}
\caption{Same as Table 1, but $\mu^+$ decays.}
\label{tab:rates2}
\end{table}

\subsection{Direct extraction of the oscillation probabilities}
From the visible energy distributions of the events, one can extract the
oscillation probabilities. The visible energy of the events are plotted
into histograms with 10 bins in energy. The $\nue\ra\numu$ oscillation 
probability in each energy bin $i$ can be computed as
\begin{eqnarray}
{\cal P}_i(\nue\ra\numu)=\frac{N_i(ws\mu)-N^0_i(ws\mu)}{\epsilon_i(p_\mu>p_\mu^{cut})N_i^0(e)}
\end{eqnarray}
where $N_i(ws\mu)$ is the number of wrong-sign muon events in the i-th bin
of energy, $N^0_i(ws\mu)$ are the background events in the i-th bin
of energy, $\epsilon_i(p_\mu>p_\mu^{cut})$ is the efficiency of the muon
threshold cut in that bin, and $N_i^0(e)$ is the number of electron events
in the i-th bin of energy in absence of oscillations. The number of events
corresponds to the statistics obtained from $\mu^+$ decays. A similar
quantity for antineutrinos ${\cal P}_i(\bar\nue\ra\bar\numu)$ will be computed with
events coming from  $\mu^-$ decays.

Similarly, the $\numu\ra\nue$ oscillation 
probability in each energy bin $i$ can be computed as
\begin{eqnarray}
{\cal P}_i(\numu\ra\nue)=\frac{N_i(wse)-N^0_i(wse)}{\epsilon_e (1-p_{conf})
N_i^0(rs\mu)}
\end{eqnarray}
where $N_i(wse)$ is the number of wrong-sign electron events in the i-th bin
of energy, $\epsilon_e$ is the efficiency for charge discrimination,
$p_{conf}$ the charge confusion,
and $N_i^0(rs\mu)$ is the number of right sign muon events
in the i-th bin of energy in absence of oscillations. The number of events
corresponds to the statistics obtained from $\mu^-$ decays. A similar
quantity for antineutrinos ${\cal P}_i(\bar\numu\ra\bar\nue)$ will be computed with
events coming from  $\mu^+$ decays.

These binned probabilities could be combined in an actual experiment in
order to perform direct searches of the effects induced by the
$\delta$-phase.

\subsection{Direct search for T-asymmetry}

For measurements involving the discrimination of the electron charge, we
limit ourselves to the lowest energy and baseline configuration ($E_\mu =
7.5\rm\ GeV$ and $L=732\rm\ km$), since we expect the discrimination of the
electron charge to be practically possible only at these lowest energies.

The binned $\Delta_T(i)$ discriminant for neutrinos is defined as
\begin{eqnarray}
\Delta_T(i)={\cal P}_i(\numu\ra\nue)-{\cal P}_i(\nue\ra\numu)
\end{eqnarray}
and a similar discriminant $\bar\Delta_T(i)$ can be computed for antineutrinos.

These quantities are plotted for neutrinos
and antineutrinos for three values of the
$\delta$-phase ($\delta=+\pi/2$, $\delta=0$ and $\delta=-\pi/2$)
in Figure~\ref{fig:directt}.
The errors are statistical and correspond to a normalization of $10^{21}$
muon decays and a baseline of $L=732\rm\ km$. 
A 20\% electron efficiency with a charge confusion
probability of 0.1\% has been assumed.
The full curve corresponds
to the theoretical probability difference.

A nice feature of these measurements is the change of sign of the effect
with respect of the change $\delta\ra -\delta$ and also with respect
to the substitution of neutrinos by antineutrinos. These changes of sign
are clearly visible and would provide a direct, model-independent,
proof for T-violation in neutrino oscillations.

In order to cross-check the matter behavior, one can also contemplate the
$CPT$-discriminants defined as
\begin{eqnarray}
\Delta_{CPT}(i)={\cal P}_i(\numu\ra\nue)-{\cal P}_i(\bar\nue\ra\bar\numu)
\nonumber\\
\bar\Delta_{CPT}(i)={\cal P}_i(\nue\ra\numu)-{\cal P}_i(\bar\numu\ra\bar\nue)
\end{eqnarray}
These quantities are independent from the $\delta$-phase and probe only
the matter effects. The change of sign of the effect
with respect to the substitution of
neutrinos by antineutrinos is clearly visible.

It should be however noted that in the case of the $CPT$ discriminant, the
statistical power is rather low, since this measurement combines the
appearance of electrons (driven by the efficiency for detecting the
electron charge) and involves antineutrinos, which are suppressed by matter
effects. Hence, the statistical power is reduced compared to the
$T$-discriminant.

\subsection{Direct search for CP-asymmetry}

In the direct search for the CP-asymmetry, we rely only on the appearance
of wrong-sign muons. We compare in this case the two energy and baselines
options. 

\begin{figure}[tb]
\centering
\epsfig{file=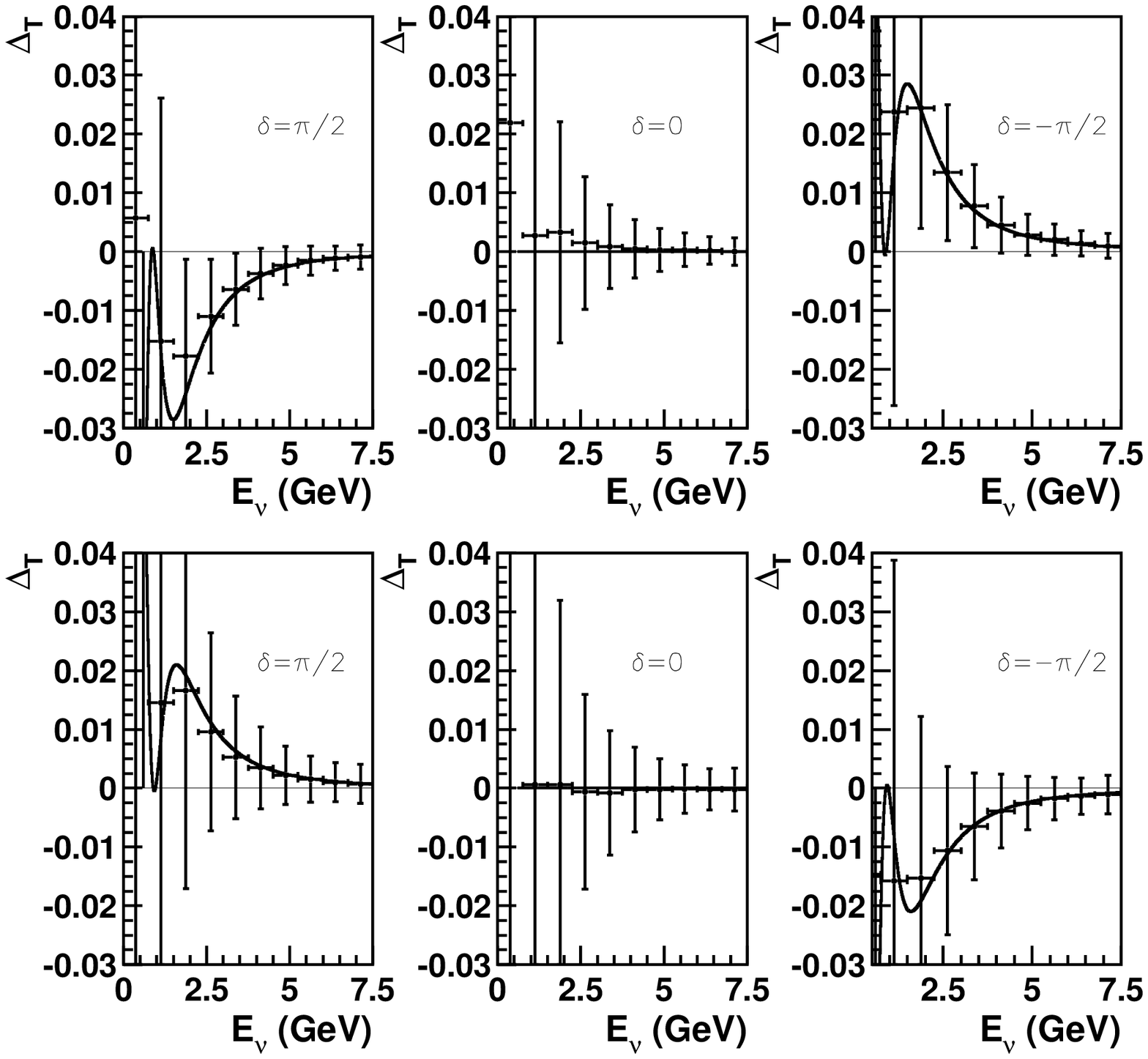,width=7.5cm}
\epsfig{file=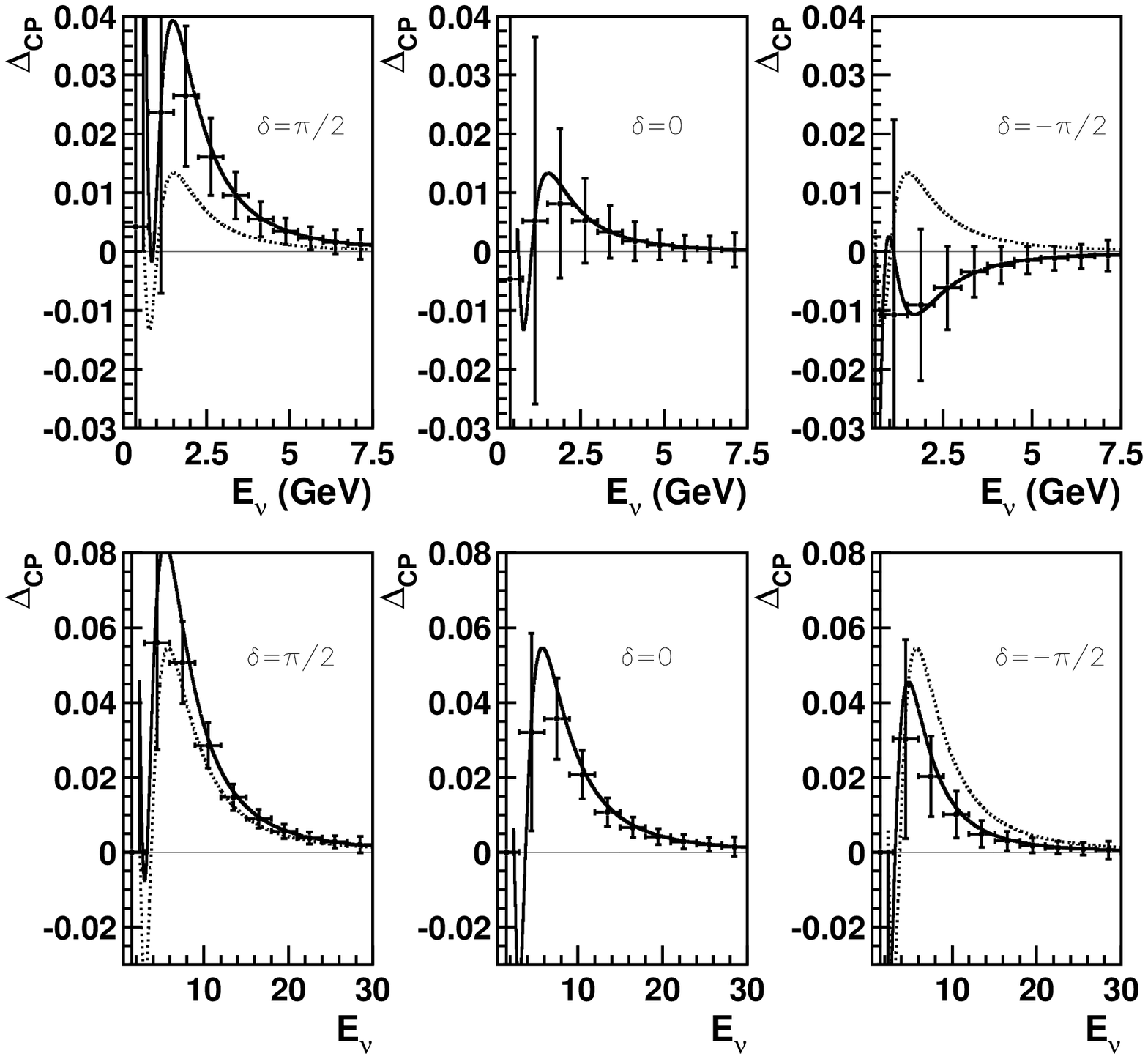,width=7.5cm}
\caption{Direct T- and CP-violation: Binned $\Delta_T(i)$(left) and 
$\Delta_{CP}(i)$(right) discriminants for neutrinos (upper
plots) and antineutrinos (lower plots) for three values of the
$\delta$-phase: $\delta=+\pi/2$, $\delta=0$ and $\delta=-\pi/2$.
The errors are statistical and correspond to the normalizations
given in the text.
A 20\% electron efficiency with a charge confusion
probability of 0.1\% has been assumed. The full curve corresponds
to the theoretical probability difference.}
\label{fig:directt}
\end{figure}

The binned $\Delta_{CP}(i)$ 
discriminant for the shortest baseline $L=732\rm\ km$, $E_\mu=7.5\rm\ GeV$
and longest baseline $L=2900\rm\ km$, $E_\mu=30\rm\ GeV$
(lower plots) for three values of the
$\delta$-phase ($\delta=+\pi/2$, $\delta=0$ and $\delta=-\pi/2$)
are shown in Figure~\ref{fig:directt}.
The errors are statistical and correspond to a normalization of
$10^{21}$($2.5\times 10^{20}$)
for $L=732(2900)\rm\ km$. The full curve corresponds
to the theoretical probability difference.
The dotted curve is the
theoretical curve for $\delta=0$ and represents the effect of 
propagation in matter.

As was already pointed out, the $\Delta_{CP}$ does not vanish even in
the case $\delta=0$, since matter traversal introduces an asymmetry. 
At the shortest baseline ($L=732\rm\ km$), these effects are rather
small. This has the advantage
that the observed asymmetry would be positive for $\delta>0$, but would
still change sign in the case $\delta\approx -\pi/2$. In the fortunate case
in which Nature has chosen such a value for the $\delta$-phase, the
observation of the negative asymmetry would be a striking sign for
CP-violation, since matter could never produce such an effect.

For other values of the $\delta$-phase, the effect is positive. It is also
always positive at the largest baseline $L=2900\rm\ km$, since at those
distances the effect induced by the $\delta$-phase is smaller than the
asymmetry introduced by the matter.

\subsection{Comparison of two methods}
The binned $\Delta_T(i)$ and $\Delta_{CP}(i)$ discriminant can be used to
calculate the $\chi^2$ significance of the effect, given the statistical
error on each bin. We compute the following $\chi^2$'s:
\begin{eqnarray}
\chi^2_T = \sum_{i}
\frac{\left(\Delta_T(i,\delta)-\Delta_T(i,\delta=0)\right)^2}
{\sigma(\Delta_T(i,\delta))^2}+
\frac{\left(\bar\Delta_T(i,\delta)-\bar\Delta_T(i,\delta=0)\right)^2}
{\sigma(\bar\Delta_T(i,\delta))^2}
\end{eqnarray}
where $\sigma(\Delta_T(i,\delta))$ is the statistical error in the bin.
Similarly, the $\chi^2$ of the CP-asymmetry is
\begin{eqnarray}
\chi^2_{CP} = \sum_{i} \frac{\left(\Delta_{CP}(i,\delta)-\Delta_{CP}(i,\delta=0)\right)^2}
{\sigma(\Delta_{CP}(i,\delta))^2}
\end{eqnarray}

We study the significance of the effect as a function of the solar mass
difference $\Delta m^2_{21}$, since the effect associated to the
$\delta$-phase will decrease with decreasing $\Delta m^2_{21}$ values. We
consider the range compatible with solar neutrino experiments,
$10^{-5}\lesssim \Delta m^2_{21}\lesssim 10^{-4}\rm\ eV^2$.

\begin{figure}[tb]
\centering
\epsfig{file=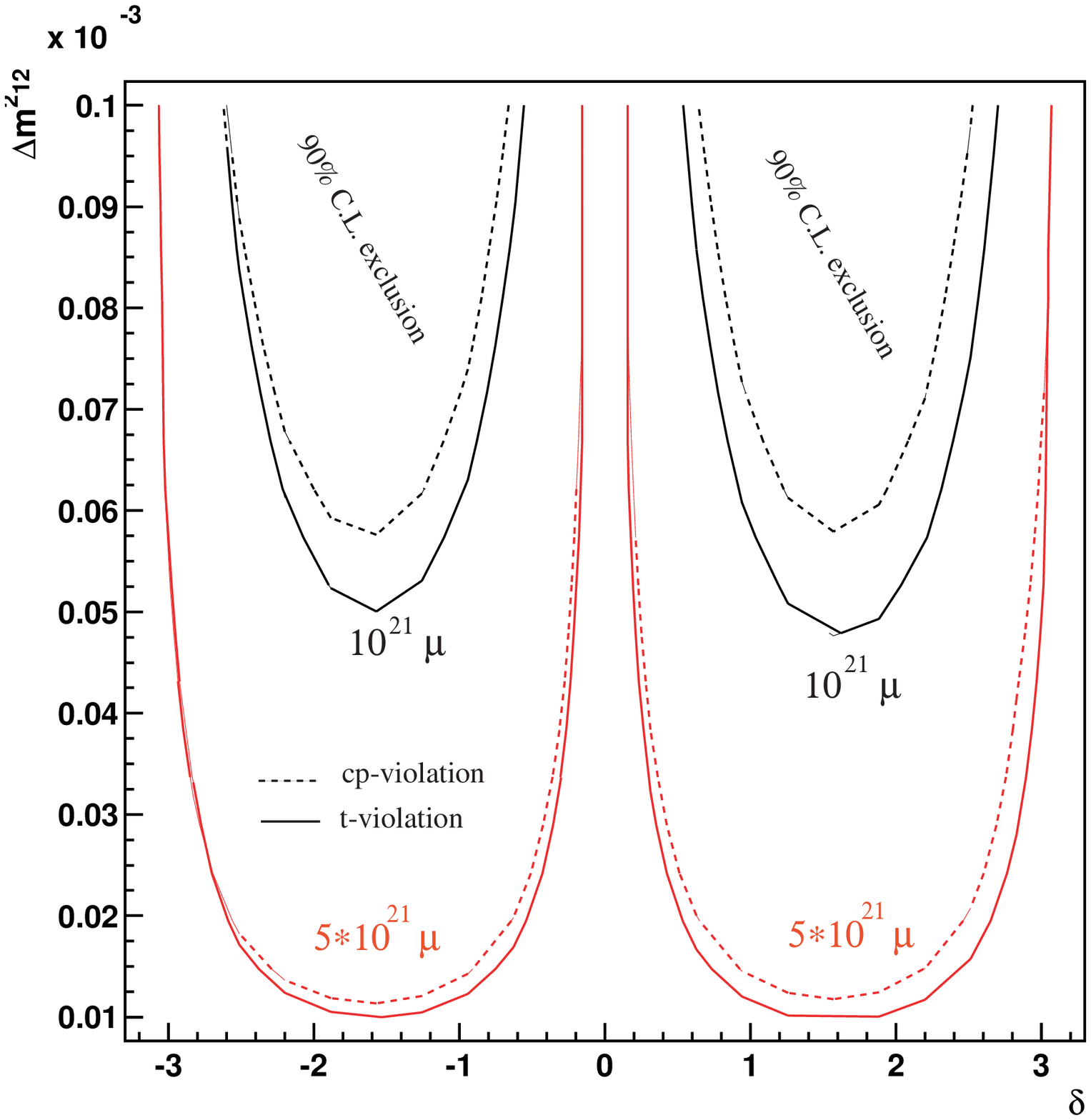,height=6cm}
\epsfig{file=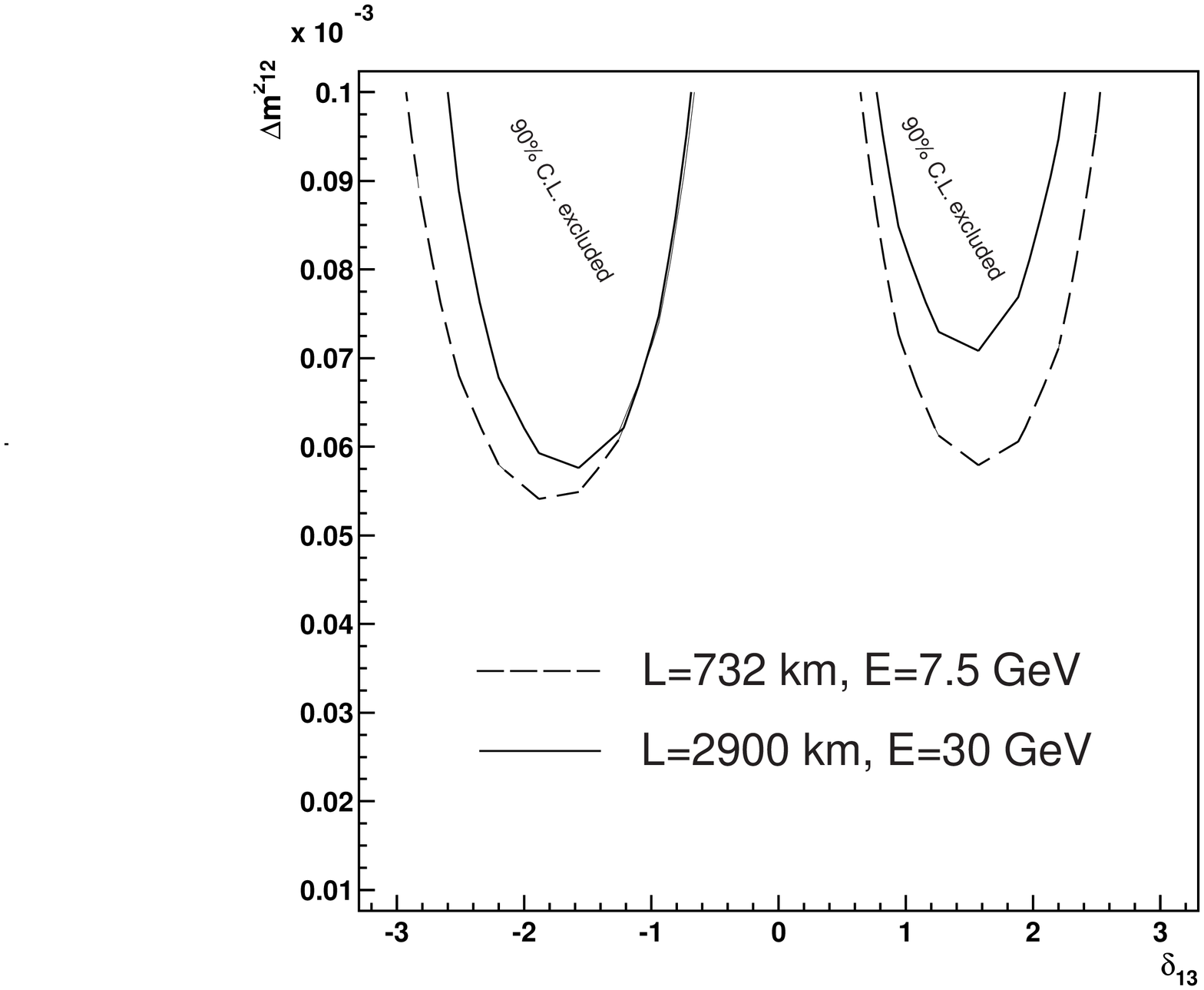,height=6cm}
\caption{Exclusion region at the 90\%C.L. in the $\delta$-phase vs $\Delta
m^2_{21}$ plane. Two regions obtained with the $\Delta_T$ and $\Delta_{CP}$
discriminants are shown. A 20\% electron efficiency with a charge confusion
probability of 0.1\% has been assumed. The normalization is
$10^{21}(5\times 10^{21})$
muon decays with energy $E_\mu=7.5\rm\ GeV$ and a baseline of $L=732\rm\
km$.}
\label{fig:excltcp}
\end{figure}

The exclusion regions obtained 
at the 90\%C.L. (defined as $\Delta \chi^2 = +1.96$)
in the $\delta$-phase vs $\Delta
m^2_{21}$ plane are shown in Figure~\ref{fig:excltcp}.
A 20\% electron efficiency with a charge confusion
probability of 0.1\% has been assumed. The normalizations assumed are
$10^{21}$ and $5\times 10^{21}$ muon decays with energy 
$E_\mu=7.5\rm\ GeV$ and a baseline of $L=732\rm\ km$.

The results are very encouraging. With $10^{21}$ muon decays, the region
$\Delta m^2_{21}\gtrsim 5\times 10^{-5}\rm\ eV^2$ is covered. For $5\times
10^{21}$ muons, this region extends down to $10^{-5}\rm\ eV^2$, in order
words, the full range of values compatible with the LMA solar data is
testable. 

If we consider that the value of $\Delta m^2_{21}$ is known and that it has a value
of $\Delta m^2_{21}=10^{-4}\rm\ eV^2$, one can constrain the values
of the $\delta$-phase within the range $|\delta|\lesssim 0.35$ or $|\delta| \gtrsim
2.8$ for $10^{21}$ muons and $|\delta|\lesssim 0.14$ for $5\times 10^{21}$ muon decays
at the 90\%C.L. We conclude that an exhaustive direct,
model-independent exploration of the $\delta$-phase, within the full range 
$10^{-5}\lesssim \Delta m^2_{21}\lesssim 10^{-4}\rm\ eV^2$ requires an
intensity of $5\times 10^{21}$ muon decays of each sign.

\section{Conclusion}

We argued that the ideal neutrino detector at the neutrino factory should
(1) have a {\it mass} in
the range of 10~kton, (2) provide {\it particle identification} to tag the flavor of
the incoming neutrino, (3) {\it lepton charge
measurement} to select the incoming neutrino helicity, 
(4) {\it good energy resolution} to reconstruct the incoming
neutrino energy, (5) possess a {\it low energy threshold}, in order to study
neutrino events in the energy region where $CP$ and $T$ effects are the cleanest and
most unambiguous (see Section 11)
and (6) {\it be isotropic} to equally well reconstruct incoming
neutrinos from different baselines (it might be more efficient to build various
sources at different baselines, than various detectors). 

A detector with
such qualities is most adapted to fully explore neutrino oscillations at the
neutrino factory. In particular, measurement of the leading muons and electrons charge is
the only way to fully simultaneously study $CP$ and $T$ violation
effects. 

We think that a magnetized liquid argon imaging detector, scaled to 10~ktons, stands
as the best choice of technique which holds the highest hope to
match the above mentioned detector requirements.

Such a detector could measure very precisely all the magnitudes of the
elements of the mixing matrix and over-constrain them, because of its
ability to reconstruct all final states including muons, electrons,
tau-like and neutral current events (see Ref.\cite{Bueno:2000fg}).

In addition, this detector
could address {\it the more challenging and most interesting
goal of the neutrino factory, which is the search for effects related to the
complex phase of the mixing matrix}. This is because it could reconstruct
the charge of both electrons and muons, and because it is perfectly adapted
to reconstruct low energy events (typ. below 15~GeV), which is the energy region
where we expect these effects to be the cleanest and the most unambiguous.

The choice of baseline and muon ring energy is in this context particularly critical.
In view of the existence of massive devices like SuperK, MINOS or ICARUS, 
it is also worth considering if these detectors at their current baselines could 
be reused in the context of
the neutrino factory. If new sites have to be found in order
to satisfy the requirements of longer baselines, major new ``investments''
will be required.

We find that the most favorable choice of neutrino energy $E_\nu$ and
baseline $L$ is in the region of the ``first maximum'' given by 
$(L/E_\nu)^{max}\simeq 400$ for $|\Delta m^2_{32}|=3\times 10^{-3}\rm\
eV^2$. 
This yields $E^{max}_{\nu}\simeq 2\rm\ GeV$ at 730~km,
$E^{max}_{\nu}\simeq 8\rm\ GeV$ at 2900~km and
$E^{max}_{\nu}\simeq 20\rm\ GeV$ at 7400~km
for $\Delta m^2_{32}=3\times 10^{-3}\rm\ eV^2$.

We showed that the discriminants of the $\delta$-phase all
scale with $L/E_\nu$. This property leaves flexibility in the choice of the actual neutrino
energy and the baseline, since only their ratio $L/E_\nu$ is
determinant.  Because of the rising neutrino cross-section with energy,
it will be more favorable to go to higher energies if the
neutrino fluence is constant. Keeping the $L/E_\nu$ ratio constant, 
this implies an optimization at longer baselines $L$. One will hence gain
with the baseline $L$ until we reach $L_{max}\approx 5000\rm\ km$ beyond
which matter effects will effectively reduce the dependence of
the oscillation probabilities on the $\delta$-phase.

As far as the neutrino energy is concerned, it is clear that the average
neutrino energy $E_\nu$ scales linearly with the muon beam energy
$E_\mu$. A non-negligible aspect of the neutrino factory is the need to
accelerate quickly the muons to the desired energy, and so, it is expected
that higher energies will be more demanding that lower ones. Eventually,
cost arguments could determine the muon energy. It could therefore be
that lower energy, more intense neutrino factories could be more advantageous
than higher energy, less intense ones. 

From the above arguments, we therefore conclude that 
a very intense neutrino factory, capable of providing more than $10^{21}$ useful 7.5~GeV muon decays, directed
towards a distance of 730~km, coupled with a magnetized liquid argon imaging
detector of about 10~kton would provide the ultimate, most comprehensive setup to study $CP$
and $T$ violation effects in neutrino flavor oscillation.

\section{Acknowledgments}
I thank Milla Baldo Ceolin for the invitation to the Neutrino Telescope Workshop
in the beautiful city of Venice. The help of M.~Campanelli and
also of A.~Bueno is greatly acknowledged.

\def\Journal#1#2#3#4{{#1} {\bf #2}, #3 (#4)}

\def\etal{{\it et\ al.}}
\def\NCA{\em Nuovo Cim.}
\def\NIM{\em Nucl. Instrum. Methods}
\def\NIMA{{\em Nucl. Instrum. Methods} A}
\def\NPB{{\em Nucl. Phys.} B}
\def\PLB{{\em Phys. Lett.}  B}
\def\PRL{\em Phys. Rev. Lett.}
\def\PRC{{\em Phys. Rev.} C}
\def\PRD{{\em Phys. Rev.} D}
\def\ZPC{{\em Z. Phys.} C}
\def\ASP{{\em Astrop. Phys.}}
\def\JETP{{\em JETP Lett.\ }}


\begin{thebibliography}{000}

\bibitem{k2k}
``E362 Proposal for a long baseline neutrino oscillation experiment,
using KEK-PS and Super-Kamiokande'', February 1995. \\
H. W. Sobel, proceedings of Eighth International Workshop on Neutrino 
Telescopes, Venice 1999, vol 1 pg 351.

\bibitem{minos} 
E.~Ables {\it et al.} [MINOS Collaboration],
``P-875: A Long baseline neutrino oscillation experiment at Fermilab,"
FERMILAB-PROPOSAL-P-875.

\bibitem{opera} A. Ereditato, K. Niwa and P. Strolin, INFN/AE-97/06
and Nagoya DPNU-97-07, 27 January 1997, unpublished.

\bibitem{icarus} ICARUS Collaboration, Laboratori Nazionali del Gran Sasso
(LNGS) Int. Note, LNGS - 94/99 (Vols I-II), unpublished; LNGS 95/10, unpublished.
P. Benetti \etal, \Journal{\NIMA}{327}{173}{1993};\Journal{\NIMA}{332}{395}{1993};
P. Cennini \etal, \Journal{\NIMA}{333}{567}{1993};\Journal{\NIMA}{345}{230}{1994};
\Journal{\NIMA}{355}{660}{1995}.

\bibitem{Arneodo:2001tx}
F.~Arneodo {\it et al.}  [ICARUS collaboration],
hep-ex/0103008.

\bibitem{pontecorvo}
B.~Pontecorvo, {\em J. Expt. Theor. Phys.} \textbf{33}, 549 (1957)
[Sov. Phys. JETP \textbf{6}, 429 (1958)];
B.~Pontecorvo, {\em J. Expt. Theor. Phys.} \textbf{34}, 247 (1958)
[Sov. Phys. JETP \textbf{7}, 172 (1958)];
Z.~Maki, M.~Nakagawa and S.~Sakata,
{\em Prog.\ Theor.\ Phys.} {\bf 28} (1962) 870;
B. Pontecorvo, {\em J. Expt. Theor. Phys} {\bf 53} (1967) 1717;
V.~Gribov and B.~Pontecorvo, Phys. Lett. B \textbf{28}, 493 (1969).

\bibitem{atmevid}
Y. Fukuda \textit{et al.} (Kamiokande Collaboration), Phys. Lett. B \textbf{335},
237
  (1994);
R. Becker-Szendy \textit{et al.} (IMB Collaboration), Nucl. Phys. B (Proc. Suppl.)
  \textbf{38}, 331 (1995);
S. Fukuda \textit{et al.} (Super-Kamiokande Collaboration), hep-ex/0009001; T.
Kajita
  (Super-Kamiokande Collaboration), Talk presented at NOW2000, Otranto, Italy,
  September 2000 (http://{\-}www.ba.infn.it/{\-}\~{}now2000);
W.W.M. Allison \textit{et al.} (Soudan 2 Collaboration), Phys. Lett. B
\textbf{449},
  137 (1999);
F. Ronga (MACRO Collaboration), Talk presented at NOW2000, Otranto, Italy,
September 2000 (http://{\-}www.ba.infn.it/{\-}\~{}now2000).

\bibitem{geers}
S. Geer, \Journal{\PRD}{57}{1998}{6989}

\bibitem{nufacwww}
Information on the neutrino factory studies and mu collider collaboration at BNL
can be found at {\it http://www.cap.bnl.gov/mumu/}.
Information on the neutrino factory studies at FNAL
can be found at {\it http://www.fnal.gov/projects/muon\_collider/}.
Information on the neutrino factory studies at CERN
can be found at {\it http://muonstoragerings.cern.ch/Welcome.html/}.

\bibitem{nufacfis}
A. de R\'ujula, M. B. Gavela and P. Hern\'andez, 
\Journal{\NPB}{547}{1999}{21}; 
V. Barger, S. Geer, R. Raja and K. Whisnant, 
Phys.\ Rev.\  {\bf D62}, 013004 (2000)
[hep-ph/9911524];
Phys.\ Rev.\  {\bf D62}, 073002 (2000)
[hep-ph/0003184]; 
A. Bueno, M. Campanelli and A. Rubbia, Nucl.\ Phys.\  {\bf B573} (2000) 27;
V. Barger, S. Geer and K. Whisnant, \Journal{\PRD}{61}{2000}{053004}; 
M. Freund, M. Lindner, S. T. Petcov and A. Romanimo, Nucl.\ Phys.\  {\bf B578}, 27 (2000)
[hep-ph/9912457];
A. Cervera \etal, Nucl.\ Phys.\  {\bf B579}, 17 (2000) [hep-ph/0002108].

\bibitem{Bueno:2000fg}
A.~Bueno, M.~Campanelli and A.~Rubbia,
Nucl.\ Phys.\ B {\bf 589}, 577 (2000)
[hep-ph/0005007].

\bibitem{Arafune:1997hd}
J.~Arafune, M.~Koike and J.~Sato,
Phys.\ Rev.\ D {\bf 56}, 3093 (1997)
[hep-ph/9703351].

\bibitem{Sielaff:2001md}
J.~Sielaff  [DONUT Collaboration],
hep-ex/0105042.

\bibitem{Jung:1999jq}
C.~K.~Jung,
Feasibility of a next generation underground water Cherenkov detector:  UNO,
NNN99 at Stony Brook, hep-ex/0005046.

\bibitem{cr} C.~Rubbia, private communication.

\bibitem{sergiam} F.~Sergiampietri, NUFACT'01, Tsukuba (Japan), May 2001.

\end{thebibliography}
\end{document}